\documentclass[prc,aps,floatfix,showpacs,twocolumn]{revtex4-2}
\usepackage{dcolumn}
\usepackage{slashed}
\usepackage{bm}
\usepackage{color}
\usepackage{graphicx}
\usepackage{pstricks}
\usepackage{amssymb,amsmath}
\usepackage[utf8]{inputenc} 							
\newcommand{\unit}{\textbf{1}}

\def\deu{{{}^2{\rm H}}}
\def\he{{{}^4{\rm He}}}
\def\heq{{{}^4{\rm He}}}
\def\a{\alpha}

\def\bmr{{\bm r}}
\def\bmv{{\bm v}}
\def\bmV{{\bm V}}
\def\bmx{{\bm x}}

\def\bmp{{\bm p}}
\def\bmk{{\bm k}}
\def\bmq{{\bm q}}
\def\bmR{{\bm R}}
\def\bmJ{{\bm J}}
\def\bmK{{\bm K}}
\def\bmQ{{\bm Q}}

\def\bmP{{\bm P}}

\newcommand{\bmsi}{{\bm \sigma}}

\newcommand{\bmta}{{\bm \tau}}

\begin{document}
\title{Dark matter scattering off $\deu$ and $\he$ nuclei within chiral effective field theory}
\author{Elena Filandri$^{\,{\rm 1},\,{\rm 2}}$ and Michele Viviani$^{\,{\rm 2}}$}
\affiliation{
$^{\rm 1}$\footnotesize{Università di Pisa, Largo Bruno Pontecorvo 3, 56127 Pisa, Italy}\\
$^{\rm 2}$\footnotesize{INFN Sezione Pisa, Largo Bruno Pontecorvo 3, 56127 Pisa, Italy}}

\date{\today}

\begin{abstract}
We study dark matter, assumed to be composed of weak interacting massive particles (WIMPs), scattering off $\deu$ and $\he$ nuclei. In order to parameterize the WIMP-nucleon interaction, the chiral effective field theory approach is used. Considering only interactions invariant under parity, charge conjugation and time reversal, we examine five interaction types: scalar, pseudoscalar, vector, axial and tensor. Scattering amplitudes between two nucleons and a WIMP are determined up to second order of chiral perturbation theory. We  apply this program to calculate the interaction rate as a function of the WIMP mass and of the magnitude of the WIMP-quark coupling constants. From our study, we conclude that the scalar nuclear response functions result much greater than the others due to their large combination of low energy constants. We verify that the leading order contributions are dominant in these low energy processes. We also provide an estimate for the background due to atmospheric neutrinos.
\end{abstract}

\maketitle

\section{Introduction}
\label{sec:intro}
A great number of gravitational anomalies have been detected since the $1930$s at galactic scales and beyond~\cite{bertone}. These anomalies, which cannot be described by the standard cosmological model, suggest the existence of a new type of particles whose properties are still unknown, forming the so-called dark matter (DM). Nowadays, one  of  the  most  important  challenges  of  physics  is  to  understand  the  nature of DM in the Universe. The long-held hypothesis is that most DM is cold and  made up of some massive particles. The leading candidates of such particles are the weakly interacting massive particles (WIMPs), still  a viable and highly motivated possibility nowadays~\cite{Feng23,Bottaro22}. Another class of candidates with the requisite thermal relic density are the so-called feebly interacting light particles~\cite{FIP23}.
While not WIMPs, feebly interacting particles with  couplings of $g\ll1$, and masses of $m\ll1$ GeV might be thought of as variations of the WIMP paradigm.

Our purpose is to study the nuclear response to WIMP scattering, assumed to be Dirac particles. This response is needed to analyze the results of various direct detection experiments, which are currently attempting to detect DM in experimental laboratories all over the world~\cite{Baudis,Cooley}.  Clearly, WIMPS are also searched for in many experiments performed in high-energy colliders, such as Tevatron and LHC, see, for example, Refs.~\cite{Goodman_2010,Goodman_2011} for the constraints imposed on the various models of DM.
The WIMPs can be assumed to be non-relativistic, since, in order to be gravitationally bound in the galaxy halos, their velocity needs to be below  about  $600$ km/s. The  typical WIMP velocity  in  the  halo  is  thus $|\frac{v_{\chi}}{c}|\approx10^{-3}$. The maximal recoil momentum transfer depends on the reduced mass of the WIMP-nucleus system and  on  the  range  of recoil  energies, $E_R$,  that  the experiments  are  measuring. This recoil energy is usually in the range of a few keV to few tens of keV~\cite{relicdensity}, while the heaviest nuclei
have masses of $m_A\approx 200$ GeV. This gives a maximal momentum transfer of $q_{max}\leq 200 \mathrm{MeV}$. This is also a typical size of the momenta exchanged between the nucleons bound inside the nucleus. Therefore, the nucleons remain non-relativistic also after scattering and the nucleus does not break apart.

In order to describe this type of scattering, the chiral effective field theory ($\chi$EFT) approach to nuclear dynamics can be used~\cite{Weinberg68,GL84,Scherer86,Weinberg90,Bernard95}. The Lagrangian interaction terms are obtained by placing the WIMP as an external source in the QCD Lagrangian~\cite{Klos13,Bishara17,Gazda17,Korber17,Fieguth18,Andreoli19,Vries23}. In this way, we can define the external currents in the $SU(2)$ flavor space and, to take into account some particular quark coupling of isoscalar-axial nature, we extend the discussion in the flavor SU(3) space~\cite{Scherer86}. Considering only interactions invariant under parity, charge conjugation and time reversal, we  examine all possible WIMP and quarks interaction types: scalar, pseudoscalar, vector, axial and tensorial. Then, within the framework of $\chi$EFT, for each interaction type,  the interaction vertices between nucleons, pions, and WIMP can be obtained~\cite{Bishara17}. From the vertices, we derive the amplitudes for nucleus-WIMP scattering by taking contributions up to next-to-next-to-leading order (N2LO) in the chiral perturbation theory ($\chi$PT). We applied this program to study the $\deu$ and $\he$ responses to the external sources and finally to calculate the interaction rate as function of the WIMP mass and of the WIMP-quark coupling constants (quantities clearly unknown). Theoretically, light nuclei are great testing laboratories as they can be described from first principles to high accuracy, moreover helium isotopes are potential experimental targets as they are sensitive to relatively light dark matter particles (mass $\leq10$ GeV)~\cite{heliumdm}.

Alternative approaches also used in the literature are Pionless effective field theory  \cite{Richardson_2022} and Galilean effective field theory (GEFT)~\cite{Fitzpatrick13,Anand14} frameworks, where all possible contact WIMP-nucleon interactions allowed by the non-relativistic symmetries are taken into account. Several calculations within GEFT approach have been already reported, setting constraints on WIMP-nuclei coupling and cross-sections~\cite{Agnes20,Akerib21,Jeong22,Heimsoth23}. Here, as stated above, we derive the interaction using $\chi$EFT. In this way, a more direct connection to WIMP-quark couplings is achieved, and in addition, thanks to chiral power counting, a direct hierarchy between the various operators contributing to the current can be established. Moreover, as shown in Ref.~\cite{hoferichter2015ch}, the relations between $\chi$EFT and GEFT operators show that the latter are not independent due to QCD effects.
With respect to other calculations of DM scattering off light-nuclei using $\chi$EFT~\cite{Korber17,Andreoli19,Vries23}, we have treated systematically the contributions of one body and two body currents for five different interactions, providing a quantitative estimate of the rate of the process in each case. In this framework, there are already calculations for Argon and Xenon, see, for example, Ref.~\cite{Aprile23}. Using $\chi$EFT we also provide an estimate for the rate of nuclear recoils induced by atmospheric neutrinos, which represents a background process.  

This paper will be organized as follows. In Section~\ref{sec:CHEFT} we will introduce the $\chi$EFT framework and define the quark-WIMP currents.
In Section~\ref{sec:WNint} we will use the $\chi$EFT to parametrize the WIMP-nucleon interaction and compute the transition amplitudes using the $\chi$PT.
In Section~\ref{sec:rate} the interaction rate between nucleus and WIMP will be calculated using the multipolar expansion of the currents.
In Section~\ref{sec:res} we will present the results, in particular the nuclear responses to WIMP scattering and the interaction rate for each interaction type.
Finally, in Section~\ref{sec:conc} we will discuss the conclusions and perspectives of the present work.

\section{Chiral effective field theory framework and definition of external currents}
\label{sec:CHEFT}

The $\chi$EFT treatment of a general WIMP interaction has been developed and employed in calculations in several works~\cite{Klos13,Bishara17,Gazda17,Korber17,Fieguth18,Andreoli19,Vries23},
usually in the heavy-barion approach. Here we will start from the relativistic chiral Lagrangian and then make a non-relativistic expansion of the final amplitudes~\cite{baroni15}.

We start from the following general Lagrangian at $\approx 1$ GeV energy scale,
\begin{eqnarray}\label{eq:L0}
  \mathcal{L}^{\rm q}_{\rm QCD}&=&\mathcal{L}_{\rm QCD}^{{\cal M}=0}\nonumber\\
  && +\bar{q}(x)\gamma^\mu\big(v_\mu(x)+\frac{1}{3}v_\mu^{(s)}(x)+\gamma^5a_\mu(x)\big)q(x)\nonumber\\
  &&-\bar{q}(x)\big(s(x)-i\gamma^5p(x)\big)q(x)\nonumber\\
  && +\bar{q}(x)\sigma^{\mu\nu} t_{\mu\nu}(x)q(x) \,,
\end{eqnarray}
where
\begin{equation}
q(x)=\left(\begin{array}{c} 
         u(x)\\
         d(x)
         \end{array}\right), 
\end{equation}
$u(x)$ and $d(x)$ being the fields of the $u$ and $d$ quarks.
Above $\mathcal{L}_{\rm QCD}^{{\cal M}=0}$ is the QCD Lagrangian for massless quarks, while $s(x)$, $p(x)$,
$v(x)_\mu$ , $v_\mu^{(s)}(x)$, $a_\mu(x)$, and $t_{\mu\nu}(x)$ are external currents, to be specified below. As it is well known,
$\mathcal{L}_{\rm QCD}^{{\cal M}=0}$ is invariant under the chiral group under independent
unitary transformation of the right and left components of the quark field $q(x)$.

The Lagrangian given in Eq.~(\ref{eq:L0}) is written in such a way to be invariant under {\it local} chiral transformations~\cite{GL84}. 
In general, external source fields are Hermitian matrices in the isospin space;  the scalar (S) and pseudoscalar (P) charge densities are written explicitly as
\begin{equation}\label{sp}
s(x)=\sum_{a=0}^{3}{\tau_{a}}s^a(x), \hspace{1cm}
p(x)=\sum_{a=0}^{3}{\tau_{a}}p^a(x),
\end{equation}
where $ \tau_0 \equiv \unit $ and $\tau_i$, $i=1,3$, are the Pauli matrices. 
The vector (V), axial (A), and tensor (T) current densities are defined as
\begin{alignat}{2}
  v_\mu(x)= & \sum_{a=1}^3\frac{\tau_{a}}{2}v^a_\mu(x)\,,
  \qquad v^{(s)}_\mu(x)\equiv v^{(s)}_\mu(x)\tau_0\,, \\
  a_\mu(x)= & \sum_{a=1}^3\frac{\tau_{a}}{2}a^a_\mu(x)\,,\\
  t_{\mu\nu}(x)=& \sum_{a=0}^3\frac{\tau_{a}}{2}t^a_{\mu\nu}(x)\,.  
\end{alignat}
An eventual isoscalar external axial current $a_\mu^{(s)}(x)$ would couple to the isoscalar axial quark current,
\begin{equation}
  \bar{q}(x)\gamma_\mu\gamma^5q(x)
\end{equation}
however the latter quantity is not conserved due to the anomaly of the U(1)$_\mathrm{A}$ group~\cite{peskin}.
Therefore, it is not possible to construct an invariant Lagrangian with $a_\mu^{(s)}(x)$. This case
will be treated explicitly by considering the $\chi$EFT in the SU(3) flavor space~\cite{Scherer86}, see below. 

These external sources can be used to parametrize the coupling of quarks to electroweak field, and also to WIMPs. Moreover,  they can be used to insert explicit violations of the chiral symmetry in the Lagrangian. For example,
the explicit violation induced by the non-zero values of the quark masses can be incorporated by assuming $s(x)={\cal M}+\cdots$, where
\begin{equation}
  \mathcal{M}=\left(\begin{array}{cc} 
        m_u & 0\\
        0 & m_d
         \end{array}\right)\,,\label{eq:qmass}
\end{equation}
$m_u$ and $m_d$ being the $u$ and $d$-quark mass, respectively.

At hadronic level, it is then possible to write an effective Lagrangian involving nucleonic and pionic degrees of freedom and the various external currents and charge densities~\cite{Weinberg68,Scherer86,Weinberg90}. The symmetries used to build this Lagrangian are
{\it i)} the chiral symmetry, {\it ii)}  the Lorentz invariance and (eventually) {\it iii)} the discrete symmetries of
charge conjugation $C$ and parity $P$. With these Lagrangians it is possible to treat processes of momenta $Q\ll\Lambda_\chi$,
with $\Lambda_\chi \approx 4\pi f_{\pi}\approx 1$ GeV~\cite{georgi09}, where $f_{\pi}\simeq 92.4$ MeV is identified as the
charged pion decay constant~\cite{gross04}. If the chiral symmetry was an exact symmetry of the theory, the momentum $Q$
would be the only expansion parameter. As we have seen before, this is not true; the chiral symmetry is explicitly broken
by the mass term of the quarks that generates the mass of the pion $m_\pi$. This quantity reappears in the $\chi$EFT as a new
expansion parameter. However also $m_\pi$  is a small parameter compared to $\Lambda_\chi$, so we have two expansion
scales: $Q/\Lambda_{\chi}$ and $m_\pi/\Lambda_{\chi}$. From now, we will indicate with $Q$ both the typical momentum
and the mass of the pion. If we limit the range of $Q$ between zero and the mass difference between the baryon
$\Delta$(1232) and the nucleon, we can take as effective degrees of freedom only the pions and the nucleons,
without including heavier mesons or barions. In constructing this effective Lagrangian, a number of
coupling constants appear, the so-called low-energy constants (LECs). These
coupling constants can be fixed from experimental data, or from Lattice calculations.  The LECs entering this study
will be discussed in Subsect.~\ref{sec:lecs}. 

Let us now consider a general Lagrangian describing the interaction of quarks and the WIMP,
the latter assumed to be a Dirac fermion~\cite{Klos13},
\begin{eqnarray}\label{lqks}
  \mathcal{L_\chi}&=&\frac{1}{\Lambda^3}\sum_{f}\left[ C^{S}_{f}\bar{\chi}\chi m_f \bar{f}f+C^{P}_f\bar{\chi}i\gamma_{5}
    \chi m_f\bar{f}i\gamma_5 f\right]\nonumber \\
  &+&\frac{1}{\Lambda^2}\sum_{f}\left[ C^{V}_{f}\bar{\chi} \gamma^\mu \chi\bar{f} \gamma_\mu f
    +C^{A}_f\bar{\chi}\gamma^\mu \gamma_{5}\chi \bar{f}\gamma_\mu \gamma_5 f\right]\nonumber \\
   &+&\frac{1}{\Lambda^2}\sum_{f}\left[ C^{T}_{f}\bar{\chi}\sigma^{\mu \nu}\chi\bar{f}\sigma_{\mu \nu}f\right], 
\end{eqnarray}
where $\Lambda$ is a high energy scale, $\chi$ is the WIMP field, $f$  the field of quark of flavor $f=u, d, \ldots$ and
the Wilson coefficients $C_i$ are unknown parameters. They should in principle be fixed by choosing a particular
high energy WIMP model, thus they parameterize the effect of new physics associated with the energy scale $\Lambda$.
This scale is assumed to be very large ($\gg1$ TeV) but clearly it is also unknown. To render the scalar and
pseudoscalar matrix elements renormalization-scale invariant, the quark masses $m_f$ in the definition of
the respective operators has been explicitly included~\cite{Klos13}. Note that we have limited ourselves to consider
interactions even under parity and charge conjugation. The theory can be readily generalized to treat other cases,
as the inclusion of parity and/or charge conjugation violating terms, or the cases of either the WIMP
being a scalar or a Majorana fermion~\cite{Goodman_2011, Bishara17}. These cases will be considered in a forthcoming paper. 

Since we are interested to interaction of the WIMP with nuclei, so usually we can limit ourselves to include
in the sum in Eq.~(\ref{lqks}) only the quarks $u$ and $d$, but in case of the axial term we will include
also the quark $s$ (see below).

For the sake of simplicity, in the following we will define:
\begin{equation}
\frac{C^X_\pm}{\Lambda_S^2}=\frac{1}{\Lambda^3}\left(\frac{C^{X}_{u} m_u\pm C^{X}_{d}m_d}{2}\right)\,,\quad X=S,P
\end{equation}
and
\begin{equation}
\frac{C^X_\pm}{\Lambda_S^2}=\frac{1}{\Lambda^2}\left( \frac{C^{X}_{u}\pm C^{X}_{d}}{2}\right)\,,\quad X=V,A,T
\end{equation}
where the new parameter $\Lambda_S$ is inserted only for dimensional reasons. Hereafter, we have taken $\Lambda_S=1$ GeV. Adding
the Lagrangian $\mathcal{L_\chi}$ to the QCD Lagrangian, the resultant Lagrangian can be cast in the form given in Eq.~(\ref{eq:L0}),
where,
\begin{eqnarray}
  s(x)&=&\mathcal{M}-\frac{1}{\Lambda_S ^2}\left(C^S_++C^S_- \tau_z \right)\bar{\chi}\chi\,,\label{scurr}\\
  p(x)&=&\frac{1}{\Lambda_S ^2}\left(C^P_++C^P_- \tau_z \right)\bar{\chi}i\gamma_{5}\chi\,,\label{pcurr}\\
  \frac{1}{3}v_\mu ^s(x)&=&\frac{1}{\Lambda_S^2}C^V_+\bar{\chi} \gamma^\mu \chi\,,\label{vscurr}\\
  v^\mu(x)&=&\frac{1}{\Lambda_S^2}C^V_-\tau_z \bar{\chi} \gamma^\mu \chi\,,\label{vcurr}\\
  t^{\mu\nu}(x)&=&\frac{1}{\Lambda_S^2}\left(C^T_++C^T_-\tau_z\right) \bar{\chi}\sigma^{\mu \nu}\chi\,,\label{tcurr}
\end{eqnarray}
where in the scalar current $s(x)$ we have included also the quark mass term. 
Note that above we have not considered the axial coupling. This case will be treated in the next subsection. 

\subsection{Axial current}\label{su3ax}
Taking into account also the quark $s$, the field $q(x)$ becomes
\begin{equation}
q(x)=\left(\begin{array}{c} 
         u(x)\\
         d(x)\\
         s(x)
         \end{array}\right).
\end{equation}
Using the chiral limit (the masses of the quarks $u$, $d$ and $s$ zero), we can find a relation between one of the currents conserved in SU(3) and the isoscalar axial term. One has
\begin{equation}\label{axialA}
\langle N| \bar{u}\gamma_\mu\gamma^5u+\bar{d}\gamma_\mu \gamma^5d| N \rangle\rightarrow\langle N|A^{(8)}_\mu| N \rangle,
\end{equation}
where the current $A^{(8)}_\mu$ is~\cite{Scherer86}
\begin{equation}
A^{(8)}_\mu=\bar{u}\gamma_\mu\gamma^5u+\bar{d}\gamma_\mu \gamma^5d-2\bar{s}\gamma_\mu\gamma^5s=\sqrt{3} \bar{q}\gamma_\mu \gamma^5 \lambda_8 q,
\end{equation}
and $\lambda_8$ is the Gell-Mann matrix
\begin{equation}
\lambda_8=\frac{1}{\sqrt{3}}\begin{pmatrix}
1&0&0\\
0&1&0\\
0&0&-2
\end{pmatrix}\,.
\end{equation}
The Equation~(\ref{axialA}) is valid in the hypothesis that the content of the strange quark in the nucleon vanishes,
\begin{equation}
\langle N | \bar{s}\gamma_\mu \gamma^5 s| N \rangle=0.
\end{equation}
With these premises, we can rewrite the axial current part of the Lagrangian~(\ref{lqks}) in the SU(3) space  as,
\begin{equation}
\mathcal{L}_q^{axial}= \sum_i\alpha^A_i\bar{q}\gamma_\mu\gamma^5\lambda_i q \bar{\chi}\gamma^\mu\gamma^5\chi\,,
\end{equation}
where the constants $\alpha^A_i$ are zero except
\begin{eqnarray}
\alpha^A_3&=&\frac{1}{\Lambda^2}\left(\frac{C^{A}_{u}-C^{A}_{d}}{2}\right)\equiv \frac{C^A_-}{\Lambda_S^2}\\
\alpha^A_8&=&\sqrt{3} \frac{1}{\Lambda^2}\left(\frac{C^{A}_{u}+C^{A}_{d}}{2} \right)\equiv \sqrt{3}\frac{C^A_+}{\Lambda_S^2}
\end{eqnarray}
and the Gell-Mann matrix $\lambda_3$ is the SU(3) extension of $\tau_z$,
\begin{equation}
\lambda_3=\begin{pmatrix}
1&0&0\\
0&-1&0\\
0&0&0
\end{pmatrix}.
\end{equation}
In the following we define the SU(3) external axial current to be
\begin{equation}
  a_\mu=\sum_i a_{\mu}^{(i)}\lambda_i=\sum_i \alpha^A_i\bar{\chi}\gamma_\mu\gamma^5\chi\lambda_i\,, \label{eq:axialsu3}
\end{equation}
where only $a_{\mu}^{(3)}$ and $a_{\mu}^{(8)}$ are non-vanishing. 
\section{WIMP-Nucleon Interactions}
\label{sec:WNint}
We use the EFT framework to write the WIMP-nucleon interaction and compute the transition amplitudes using the $\chi$PT. The nucleon-WIMP interaction terms will be obtained from the WIMP-quark Lagrangian given in Eq.~(\ref{lqks}) using the standard procedure~\cite{fettes00}.

We have examined all possible WIMP-quark vertex types: scalar, pseudoscalar, vector, axial
and tensor. Now, within the framework of $\chi$EFT, for each case, the interaction vertices between
nucleon and WIMP are derived. Among them, we will take into account only the dominant ones and they will be used
to get the effective Hamiltonian. A fairly self-contained summary of this derivation is provided in
Appendix~\ref{app:WIMP-N_Int} for completeness.  Here we report a summary of the Lagrangian terms taken into account:
\begin{align}
\mathcal{L}_{int }^S\!&=- 8B_c c_1 \frac{C^S_+}{\Lambda_S^2}\bar{N}N\bar{\chi}\chi-
  4B_cc_5 \frac{C^S_-}{\Lambda_S^2}\bar{N}\tau_z N \bar{\chi}\chi\nonumber\\
  &+B_c \frac{C^S_+}{\Lambda_S^2}\bar{\chi}\chi \pi^2\,,\label{finalscalarlag}\\
\mathcal{L}_{int }^P\!&=\! 2f_\pi B_c\frac{C^P_-}{\Lambda_S ^2}\bar{\chi}i\gamma^5\chi\pi_z \nonumber\\&-2B_c\frac{C^P_+}{\Lambda_S ^2}(d_{18}+2d_{19})\bar{N}\gamma^\mu\gamma^5 N\; \partial_\mu (\bar{\chi}i\gamma^5\chi)\nonumber\\
&-2B_c\frac{C^P_-}{\Lambda_S ^2}d_{18}\bar{N}\gamma^\mu\gamma^5 \tau_z N \; \partial_\mu (\bar{\chi}i\gamma^5\chi)\,, \\
\mathcal{L}_{int}^V&=i\bar{N}\gamma^\mu( \Gamma_\mu-i v^{(s)}_\mu)N \nonumber\\
  &+\frac{\mathit{c}_6}{8M}\bar{N} \sigma^{\mu\nu} F_{\mu\nu}^{+}N+\frac{\mathit{c}_7}{4M}\bar{N} \sigma^{\mu\nu}
  F_{\mu\nu}^{(s)} N \nonumber \\
  &+\frac{f_\pi^2}{2} \langle \partial_\mu U^\dag ( i U v_\mu-i v_\mu U)\rangle ,
\label{eq:LintV}\\
\mathcal{L}^A_{int}&=
  (D+F)\bar{N}\gamma^\mu\gamma^5\tau_zN a_\mu^{(3)}+
  (3F-D) \bar{N}\gamma^\mu\gamma^5N {a_\mu^{(8)}\over\sqrt{3}} \nonumber\\
   &-2f_\pi \partial^\mu \pi_z   a_\mu^{(3)}
   +{1\over f_\pi} \bar{N} \gamma^\mu (\vec\tau\times\vec\pi)_z N  a_\mu^{(3)}
   \,,\label{eq:axlag}\\
 \mathcal{L}^{T}_{int}&=\bar{N}\sigma_{\mu\nu}\frac{1}{\Lambda_S^2}
  \left(4\tilde{c}_1C^T_++2\tilde{c}_2C^T_-\tau_z\right) N \bar{\chi}\sigma^{\mu\nu}\chi,
\label{eq:telag}
\end{align}
where  $N(x)$ is the iso-doublet of nucleon fields, $\vec\pi(x)$ the triplet of pion fields, $\chi$  the WIMP field, $
  \Gamma_\mu ={1\over 2} \big[ u^\dag \partial_\mu u + u \partial_\mu u^\dag -i u^\dag v_\mu u - i u v_\mu u^\dag\big]$,
  $F^{(s)}_{\mu\nu} =\partial_\mu v^{(s)}_\nu  - \partial_\nu v^{(s)}_\mu$, $F_{\mu\nu}=\partial_\mu v_\nu  - \partial_\nu v_\mu$, 
$M$  the nucleon mass, $f_\pi$  the pion decay constant and $B_c,\,c_1,\,c_5,\,c_6,\,c_7,\,d_{18},\,d_{19},\,F,\,D,\,\tilde{c}_1,\,\tilde{c}_2$ are LECs.

Then, the amplitude for the elastic scattering of a WIMP by a two nucleon system is obtained using the time-ordered perturbation theory (TOPT) method~\cite{baroni15}.
It is given as a sum of TOPT diagrams. 
Finally, we will make a non-relativistic expansion of the amplitude in power of $Q/M \approx Q/\Lambda_\chi$. 
Using the naive counting rule, each term will be characterized by a chiral "order" $Q^\nu$, where $\nu$ is an integer number. 
The terms with the lowest value of $\nu=\nu_{min}$ are denoted as the leading order (LO) terms, 
those with $\nu=\nu_{min}+1$ as the next-to-leading order (NLO) terms, etc.
In this study, we will consider contributions up to N2LO.

The amplitude for the scatter of a WIMP by a two-nucleon system has the following general form,
\begin{eqnarray}\label{eq:JJL}
  T_{fi}&=&\bigg\{{1\over\Omega}\bigg(J_{\a_1,\a_1'}^{(1)}\delta_{\bmp_1'+\bmk',\bmp_1+\bmk}\delta_{\a_2',\a_2} \nonumber\\
  &+&J_{\a_2,\a_2'}^{(1)}\delta_{\bmp_2'+\bmk',\bmp_2+\bmk}\delta_{\a_1',\a_1}\bigg)\nonumber\\
  &+&{1\over\Omega^2}J_{\a_1,\a_1',\a_2,\a_2'}^{(2)}\delta_{\bmk_1+\bmk_2,\bmk-\bmk'}\bigg\}\cdot L_{\bmk' r',\bmk r}
\end{eqnarray}
where $\Omega$ is the normalization volume, for the sake of simplicity, in the following we will take $\Omega=1$,  and  $\a_i\equiv\{\bmp_i,s_i,t_i\}$ indicate the state of nucleon $i$ ($s_i$ and $t_i$ are the $z$-projection of the spin and isospin, respectively). Here the initial (final) state of the WIMP is specified by a 
momentum $\bmk$ ($\bmk'$) and spin projection $r$ ($r'$). The initial (final) state of the nucleon $i$ is identified
by the quantum numbers $\a_i$ ($\a_i'$). The mass of the pion, the nucleon, and the WIMP will be denoted as $m_\pi$, $M$, and $M_\chi$, respectively. 
Moreover, we define $\bmk_i=\bmp'_i-\bmp_i$, $\bmK_i=(\bmp_i+\bmp_i')/2$,
$\bmq=\bmk-\bmk'$, and $\bmQ=(\bmk+\bmk')/2$. Clearly $\bmq$ is the momentum transferred by the WIMP to the two nucleon
systems. In the following $J^{(1)}$ ($J^{(2)}$) is the so-called one-body (two-body) current, while
$L$ is the so-called WIMP current. To determine eventual three-body transition currents,
one should consider the interaction of the WIMP with a three-nucleon system. However, we will neglect this
latter contribution. Note that for the vector and axial interaction $J^{(1,2)}$ and $L$ are four vectors,
so in Eq.~(\ref{eq:JJL}) $J^{(1,2)}\cdot L\equiv J^{(1,2)}_\mu L^\mu$, while in the tensor case all quantities are
four tensors, so  $J^{(1,2)}\cdot L\equiv J^{(1,2)}_{\mu,\nu} L^{\mu,\nu}$, etc. 
Both nuclear currents $J^{(1,2)}$ and $L$ will be constructed at N2LO, independently
of each other (except for some cases). 
We refer to the Appendix~\ref{app:WIMP-N_Int} for all details of the calculation on the five examined interaction cases and for the values of the LECs used in this work.

\section{The interaction rate}
\label{sec:rate}
Let us now calculate the cross-section for the elastic scattering between a nucleus and the WIMP.
The initial state $i$ is the state with an incoming WIMP of momentum $\bmk$ and nucleus at rest in the laboratory.
The energy of this initial state is 
\begin{equation}
E_i=M_A+M_\chi+\frac{k^2}{2M_\chi},
\end{equation}
where $M_A$ is the nucleus mass, $M_\chi$ the WIMP mass and $k=|\bmk|$ is the absolute value of the initial WIMP momentum.

The final state $f$ has energy,
\begin{equation}
E_f=M_A+\frac{P'^{2}_{A}}{2M_A}+M_\chi+\frac{k'^2}{2M_\chi},
\end{equation}
where we have indicated with $k'=|\bmk'|$ the absolute value of the final WIMP momentum and with $P'_A=|\bmP'_A|$ that of the nucleus. 
The non-polarized cross-section for this process is calculated from Fermi golden rule, by mediating
over the initial polarizations and summing over the final ones,
\begin{eqnarray}
  \sigma_{fi}  &=&\frac{2\pi}{2(2J_A+1)}\sum_{r'r}\sum_{s'_A s_A}\sum_{\bmk'}
  \sum_{\bmP'_{A}}\frac{1}{v} \delta_{\bmk,\bmP'_{A}+\bmk'}\nonumber\\
  && |T_{fi}|^2 \delta\left(\frac{k^2}{2M_\chi}-\frac{P'^{2}_{A}}{2M_A}-\frac{k'^2}{2M_\chi}\right)
   \,,\label{sig1}
\end{eqnarray}
where $J_A$ is the spin of the target nuclei and the
Kronecker delta in Eq.~(\ref{sig1}) fixes the final momentum of the WIMP to be $\bmk'=\bmk-\bm{P}'_A$,
thus eliminating the sum over $\bmk'$. Moreover, $\bmv=\bmk/M_\chi$ is the velocity of the
incoming WIMP. 

We are going to compute the matrix elements $T_{fi}$ using the nucleus wave function $\Psi_A^{J_A,s_A}$ calculated in $r$-space
($s_A$ is the $z$-component of the nuclear spin, which can assume the values $-J_A,\ldots,+J_A$).
For that reason, we need to express the currents in configuration space. In general, we can write
\begin{eqnarray}
  T_{fi}&\!\!\!=\!\!\!&\sum_{X=S,P,V,A,T} \sum_{a=\pm} {C^X_a\over\Lambda_S^2} \nonumber\\
  &\!\!\! \times\!\!\!& \int e^{i\bmq\cdot\bmx} \langle \Psi_A^{J_A,s_A'}| J^{X_a}_c(\bmx)|\Psi_A^{J_A,s_A}\rangle (L^{X_a})^c d\bmx,\,\,
\end{eqnarray}
where we have put in evidence the WIMP coupling constants. Above $c$ is an index which runs over the
(eventual) Lorentz indices of the currents for the case $X$, namely it takes into account if the currents are scalar, four-vectors,
or four-tensor quantities. The $r$-space currents $J^{X_a}_c(\bmx)$ can be expressed as,
\begin{eqnarray}
  J^{X_a}_c(\bmx)&=&\sum_{i=1,A} J^{X_a,(1)}_c (i) \delta(\bmx-\tilde{\bmr}_i)\nonumber\\
  && +\sum_{i<j} J^{X_a,(2)}_c(i,j) \delta(\bmx-\tilde\bmR_{ij})
  \,,\label{eq:Jr}
\end{eqnarray}
where here the indexes $i$ and $j$ runs over the nucleons and $\tilde\bmR_{ij}=(\tilde\bmr_i+\tilde\bmr_j)/2$. Note that in the previous
expression $\tilde\bmr_i\equiv\bmr_i-\bmR_{CM}$, where $\bmR_{CM}$ is the position of the nucleus center-of-mass (CM).
The quantities $J^{X_a,(1)}_c$ and $J^{X_a,(2)}_c$ are written in terms of operators which act
on the nucleonic degrees of freedom (as $\tilde\bmr_i$, $\sigma_i$, $\nabla_i$, etc).
The dependence on $\bmR_{CM}$  has been already integrated out to obtain the 
momentum conservation delta in Eq.~(\ref{sig1}).
The quantities $J^{X_a,(1)}_i$ and $J^{X_a,(2)}_{ij}$  are related to the Fourier transforms of the one- and two-body
currents described in Appendix~\ref{app:WIMP-N_Int}. These Fourier transforms
are obtained without applying any cutoff in the momentum integrals. Note that the
WIMP currents $(L^{X_a})^c$ are exactly those reported in Appendix~\ref{app:WIMP-N_Int}.
It is convenient now to perform a multipolar expansion of the matrix elements. For simplicity, hereafter we will
concentrate in the case where only a single coupling constant $C^X_a$ is different from zero.
For the vector and axial cases, the multipolar expression is given by
\begin{alignat}{1}\label{MEinM}
  \big<&\Psi_A^{J_A,s_A'}\big|\int e^{i\bmq\cdot \bmx} J^{X_a}_\mu(\bmx)(L^{X_a})^\mu d\bmx\big| \Psi_A^{J_A,s_A}\big> =\nonumber\\
  &=(-1)^{J_A-s_A}\bigg(\sum_{l\geq 0}^{\infty}\sum_{m=-l}^{l}i^l\mathcal{D}^{l}_{m, 0}(\varphi,\theta,-\varphi)\sqrt{4\pi}\nonumber\\
  &\quad \times(J_A s'_A J_A -s_A|l m)\big\{L_0 X^C_{l}-L_z X^L_{l}\big\}\nonumber\\
  &\quad- \sum_{l=1}^{\infty}\sum_{m=-l}^{l}\sum_{\lambda=\pm1}i^l\mathcal{D}^{l}_{m, \lambda}(\varphi,\theta,-\varphi)\sqrt{2\pi}\nonumber\\
  &\quad\times (J_A s'_A J_A-s_A|l m) L_{-\lambda}\left\{\lambda X^M_{l}+X^E_{l}\right\}\bigg)\,, 
\end{alignat}
where $X^C_{l}$, $X^L_{l}$, $X^E_{l}$, and $X^M_{l}$ are the charge, longitudinal, electric and magnetic
reduced matrix elements (RMEs), respectively. Above $L_z$, $L_{\pm1}$ are defined with respect to a reference system with $\hat z=\hat \bmq$. The corresponding expressions for the scalar and pseudoscalar cases
are obtained by retaining the charge RMEs only, while for the tensor case the longitudinal, electric and magnetic RMEs only.
Above $\theta$ and $\varphi$ are the spherical angles of $\bmq$ with respect to the laboratory system (to be specified later).

Now, using the following properties
\begin{align}
&\sum_{s_A s'_A}(J_A s'_A J_A-s_A|l m)(J_A s'_A J_A-s_A|l' m')=\delta_{l'l}\delta_{m'm}\,,\\
&\sum_m\mathcal{D}^{l}_{m, \lambda}{\mathcal{D}^{l}}^*_{m, \lambda'}=\delta_{\lambda,\lambda'}\,,
\end{align}
we obtain
\begin{alignat}{1}
  d&\sigma_{fi}=
  \frac{\pi}{2J_A+1}\frac{(C^X_a)^2}{\Lambda_S^4}\sum_{r'r}\sum_{\bmP'_{A}}\delta\bigg(\frac{\bmk\cdot \bmP'_A}{M_\chi}
  -\frac{\bmP'_A{}^2}{2\mu}\bigg)\frac{1}{v}\nonumber\\
  &\times\bigg\{(4\pi)\sum_{l\geq 0}\bigg[L_0L_0{}^* |X^C_{l}|^2+L_z L_z{}^*| X^L_{l}|^2\nonumber\\
  &\qquad \qquad -2 L_0{L_z}^* Re\big(X^C_{l}X^L_{l}{}^*\big)\bigg]\nonumber\\
  &+(4\pi)\sum_{l\geq1}L_{1} L_{1}^*\big(|X^M_l|^2+|X^E_l|^2\big)\bigg\}\,,
\label{sigma2}
\end{alignat}
where we used the fact that  $L_{0}L^*_{z}=L_{0}^* L_{z}$ and  $L_{-1}L^*_{-1}=L_{+1}L_{+1}^*$. Above $\mu$ is the reduced mass of WIMP-nucleus system. 

The RMEs are calculated evaluating the matrix elements in a coordinate system where $\bmq$ is along $z$
(so that $\theta=\varphi=0$), and then reversing Eq.~(\ref{MEinM}).
Once the various RMEs have been determined, they can be used in Eq.~(\ref{MEinM}) to obtain the matrix elements for a
generic direction of the momentum transfer $\hat{\bmq}$. From Eq.~(\ref{sigma2}), in the continuous limit
$\Omega\rightarrow\infty$ the sum on $P_A'$ transforms in an integral, then
\begin{alignat}{1}
  \frac{d^2\sigma}{dE'_Ad\hat{\bmP_A'}}&=\frac{\pi}{(2J_A+1)(2\pi)^3}\frac{(C^X_a)^2}{\Lambda_S^4}
  \sum_{r'r}M_A\,\delta\bigg(\bmv\cdot\hat \bmP'_A-\frac{P'_A}{2\mu}\bigg)\nonumber\\
  &\cdot\frac{1}{v}\bigg\{(4\pi)\sum_{l\geq 0}\bigg[L_0{L_0}^* |X^C_{l}|^2+L_z {L_z}^*| X^L_{l}|^2\nonumber\\
  &\qquad\qquad -2L_0{L_z}^* Re\big(X^C_{l}X^L_{l}{}^*\big)\bigg]\nonumber\\
  &+(4\pi)\sum_{l\geq1}L_{1} L_{1}^*\big(|X^M_{l}|^2+|X^E_{l}|^2\big)\bigg\}\,,\label{d3sigma}
\end{alignat}
where $E'_A=P_A^{\prime 2}/2M_A$ is the recoiling nucleus kinetic energy. 

The double-differential rate of interactions per second induced by the WIMP will be given by~\cite{Potforliq},
\begin{equation}\label{DDR}
   \frac{d^2R}{dE'_Ad\hat{\bmP_A'}}=N_\chi N_A \int d^3 \bmv\,v\,f(\bmv)\frac{d^2\sigma}{dE'_Ad\hat{\bmP_A'}}
\end{equation}
where $N_A$ is the number of nuclei in the target, $N_\chi$ the numerical density of WIMPs,
and $f(\bmv)$ the velocity distribution for the incoming WIMP. We assume the Standard Halo
Model (SHM)\cite{Cooley,Potforliq}, i.e. a Maxwell-Boltzmann WIMP velocity distribution of width $\sigma_v$,
\begin{equation}
   f(\bmv)=\frac{1}{\sqrt{(2\pi\sigma_v^2)^3}}e^{-\frac{1}{2}(\frac{\bmv+\bmV}{\sigma_v})^2}\,,
\end{equation}
where $\bmV$ is the earth velocity relative to Galactic center.

It is now necessary to compute the factors $\sum_{r'r}L_iL_j^*$ with $i,j=0,z,\lambda$ in Eq.~(\ref{d3sigma}).
To do this we will need to consider explicitly the form of the currents obtained for each type of interaction,
however we will give here the most general result:
\begin{equation}\label{LL}
   {1\over2} \sum_{r'r}L_iL_j^*=a+\bm{b}\cdot\bm{u}+cu^2+(\bm{d}\cdot\bm{u})^2+O(\bm{u}^3)\,,
\end{equation}
where $a$, $\bm{b}$, $c$ and $\bm{d}$ are parameters depending only on $\bmq$, $\bmV$ and $M_\chi$, and $\bm{u}=\bmv+\bmV$.
In the expression above we have used the fact that $u\approx k/M_\chi\ll1$. 
Therefore, substituting the expression given in Eq.~(\ref{LL}) in Eqs.~(\ref{d3sigma}) and~(\ref{DDR}), we find that we
have to evaluate the following integrals (the so-called Radon transform)
\begin{alignat}{1}
  \!\!\!\!I&(a,\bm{b},c,\bm{d})=\nonumber\\
  &=\!\! \int\!\! d^3 \bm{u}\,\,\frac{e^{-\frac{\bm{u}^2}{2\sigma_v^2}}}{\sqrt{(2\pi\sigma_v^2)^3}}
  \bigg(\!a\!+\!\bm{b}\cdot\bm{u}\!+\!cu^2\!+\!(\bm{d}\cdot\bm{u})^2\!\bigg)\delta(\bm{u}\cdot\hat{\bmq}\!-\!A)\nonumber\\
  &=\!\! \frac{e^{-\frac{A^2}{2\sigma_v^2}}}{\sqrt{2\pi\sigma_v^2}}\bigg(\!a\!+\!\bm{b}\!\cdot\!
  \hat{\bmq} A\!+\!2c\sigma_v^2\!+\!cA^2\!+\!d^2\sigma_v^2\!-\!(\bm{d}\cdot\hat{\bmq})^2(\sigma_v^2\!-\!A^2)\!\!\bigg)\label{I}
\end{alignat}
where we used the condition $\bmP'_A=\bmq$ and called $A=\bmV\cdot\hat{\bmq}+\frac{q}{2\mu}$.

Finally, the general expression of the interaction rate when only one coupling constant $C^X_a$ is
different from zero is,
\begin{equation}
  \frac{d^2R}{dE'_Ad \hat{\bmP_A'}}=\frac{N_\chi N_A M_A}{(2J_A+1)\pi}\frac{(C^X_a)^2}{\Lambda_S^4}  \sum_{\alpha=1,4} F^X_\alpha(q)
  I^X_\alpha \,,\label{eq:d3rate}
\end{equation}
where 
\begin{eqnarray}
  F^X_1(q)&=&\sum_l |X^C_l|^2\,, \\
  F^X_2(q)&=&- \sum_l 2Re\left(X^C_l X^L_l{}^*\right)\,,\\
  F^X_3(q)&=& \sum_l|X^L_l|^2\,,\\
  F^X_4(q)&=&\sum_l\left(|X^M_l|^2+|X^E_l|^2\right)\,
\end{eqnarray}
are the nuclear structure functions and $I_\alpha$ the quantities calculated in Eq.~(\ref{I})
for each case. For the scalar and pseudoscalar cases, we can assume $F_{2,3,4}=0$, while for the tensor case, $F_1=F_2=0$.
Actually, for the tensor case, we have the contributions of currents $\bmJ_A$ and $\bmJ_B$,
the first given by $J^{ij}\equiv\epsilon_{ijl} J_A^l$ 
and the second given by $J^{0i}\equiv J_B^i$, see Eqs.~(\ref{eq:j1teij}) and ~(\ref{eq:j1te0i}). Correspondingly, two
set of RMEs are calculated, $X^{L,M,E}_l(A)$ and $X^{L,M,E}_l(B)$. The expression of the rate in this case reads:
\begin{eqnarray}
  \frac{d^2R}{dE'_A d\hat{\bmP_A'}}&=&\frac{N_\chi N_A M_A}{(2J_A+1)\pi}\frac{(C^T_a)^2}{\Lambda_S^4}\nonumber\\
  & \times& \sum_{\alpha=3,4} 4\big[ F^{T,A}_\alpha(q) I^{T,A}_\alpha + F^{T,B}_\alpha(q) I^{T,B}_\alpha\nonumber\\
  & & \qquad\qquad +2 F^{T,AB}_\alpha(q) I^{T,AB}_\alpha \big]\,,\label{eq:d3ratet}
\end{eqnarray}
where
\begin{eqnarray}
  F^{T,A}_3(q)&=& \sum_l|X^L_l(A)|^2\,,\\
  F^{T,B}_3(q)&=& \sum_l|X^L_l(B)|^2\,,\\
  F^{T,AB}_3(q)&=&0\,,\\
  F^{T,A}_4(q)&=&\sum_l\left(|X^M_l(A)|^2+|X^E_l(A)|^2\right)\,,\\
  F^{T,B}_4(q)&=&\sum_l\left(|X^M_l(B)|^2+|X^E_l(B)|^2\right)\,,\\
  F^{T,AB}_4(q)&=&\sum_l\Im\big[X^E_l(A)X^{M}_l(B)^*\nonumber\\
    &&  \qquad+X^M_l(A)X^{E}_l(B)^*\big]\,.
\end{eqnarray}
Note the extra factor $4$ in Eq.~(\ref{eq:d3ratet}), coming from the evaluation of $J^{\mu\nu}L_{\mu\nu}$
and the presence of interference terms. The expressions of all quantities $I^X_\alpha$ are reported in Appendix~\ref{app:b}.

\section{Results}
\label{sec:res}

In this Section, we report the results of the calculation of the various quantities for the
deuteron-DM and $\heq$-DM scattering.

\subsection {Deuteron-DM scattering}
\label{sec:resd}

Since the deuteron has spin $1$, then in the matrix elements $J_A=1$. Consequently, we can
have RMEs of multipoles $l=0,1,2$. However, due to the well-defined parity of the nuclear ground state and of the
multipolar transition operators, some of the multipoles vanish. In Table~\ref{tab:rmed} we report
the non-vanishing RMEs for the various cases and the various chiral orders.  The deuteron ground state
wave functions have been calculated using the Argonne V18 (AV18) potential~\cite{AV18} and a
chiral potential developed at next-to-next-to-next-to-next-to-leading order (N4LO) in Ref.~\cite{MEN17}.
There are three versions of such a potential, depending on the cutoff used to regularize it for large momenta. 
In Table~\ref{tab:rmed}, we have used the potential regularized with cutoff of $500$ MeV, hereafter denoted as the N4LO500 potential.

\begin{table*}[bth]
  \caption{\label{tab:rmed}
   The RMEs $X^C_{l}$, $X^L_{l}$, $X^E_{l}$ and $X^M_{l}$ contributing
   to DM scattering off deuterons calculated for two widely used NN  interactions, 
   the AV18~\protect\cite{AV18} and N4LO500~\protect\cite{MEN17} potentials. 
   Here $q=0.05$ fm${}^{-1}$. In the fourth column, we report the order of the transition operator.
   Since the deuteron has zero isospin, only the RMEs of the
   isoscalar operators are reported. For the S and V (P, A and T) cases, $X^C$, $X^L$
   and $X^E$ are purely real (imaginary), while $X^M$ are purely imaginary (real). 
   The notation $X\pm Y$ is a shortcut for $X\; 10^{\pm Y}$.
   The equation numbers reported in the third column 
   specify the operators from which these RMEs are calculated, as
   reported in detail in Appendix~\protect\ref{app:WIMP-N_Int}.}
  
    \begin{center}
      \begin{tabular}{llll|ccc|ccc}
        \hline\hline
         Int. & RME & Operator & order & \multicolumn{3}{c}{AV18} & \multicolumn{3}{c}{N4LO500} \\
         \hline
         S  & & & & $l=0$ & $l=1$ & $l=2$ & $l=0$ & $l=1$ & $l=2$  \\
            &  $X^C_{l}$  & Eq.~(\ref{J1scalar0}) & LO   & $-0.144+02$ & & $-0.229-02$ & $-0.144+02$  & & $-0.231-02$ \\
            &                & Eq.~(\ref{J2scalar1}) & NLO  & $-0.218+00$ & & $+0.328-03$ & $-0.806-01$  & & $+0.336-03$ \\
            &                & Eq.~(\ref{J1scalar2}) & N2LO & $+0.153+00$ & & $-0.467-05$ & $+0.103+00$  & & $-0.829-05$ \\
         \hline
         P  & & & & $l=0$ & $l=1$ & $l=2$ & $l=0$ & $l=1$ & $l=2$  \\
            &  $X^C_{l}$  & Eq.~(\ref{JeLpseudo1}) & N2LO   & & $-0.367-01$ & &  & $-0.377-01$ & \\
        \hline
         V  & & & & $l=0$ & $l=1$ & $l=2$ & $l=0$ & $l=1$ & $l=2$  \\
         &  $X^C_{l}$  & Eq.~(\ref{eq:rho1v1}) & LO   & $-0.293+01$ & &  $-0.465-03$ & $-0.293+01$ & & $-0.470-03$ \\
         &                & Eq.~(\ref{eq:rho1v3}) & N2LO & $+0.289-04$ & &  $+0.320-05$ & $+0.294-04$ & & $+0.320-05$ \\
         &  $X^L_{l}$  & Eq.~(\ref{eq:j1v1})   & NLO  & $-0.769-02$ & &  $-0.120-05$ & $-0.769-02$ & & $-0.120-05$ \\
         &  $X^M_{l}$  & Eq.~(\ref{eq:j1v1})   & NLO  &  & $-0.150-01$  & & & $-0.152-01$ &  \\         
        \hline
         A  & & & & $l=0$ & $l=1$ & $l=2$ & $l=0$ & $l=1$ & $l=2$  \\
         &  $X^C_{l}$  & Eq.~(\ref{eq:rho1a1}) & NLO  & & $+0.593-03$ & & & $+0.609-03$ & \\
         &  $X^L_{l}$  & Eq.~(\ref{eq:j1a0})   & LO   & & $+0.226+00$ & & & $+0.232+00$ & \\
         &  $X^L_{l}$  & Eq.~(\ref{eq:j1a2})   & N2LO & & $-0.469-03$ & & & $-0.781-03$ & \\
         &  $X^E_{l}$  & Eq.~(\ref{eq:j1a0})   & LO   & & $-0.319+00$ & & & $-0.328+00$ & \\
         &  $X^E_{l}$  & Eq.~(\ref{eq:j1a2})   & N2LO & & $+0.660-03$ & & & $+0.110-02$ & \\
        \hline
         T  & & & & $l=0$  & $l=1$ & $l=2$ & $l=0$ & $l=1$ & $l=2$  \\
         &  $X^L_{l}(A)$  & Eq.~(\ref{eq:jta1})   & LO   & & $-0.422+00$ & & & $-0.433+00$ & \\
         &  $X^L_{l}(B)$  & Eq.~(\ref{eq:jtb1})   & NLO  & $-0.280-02$ &  & $-0.290-03$ & $-0.285-02$ &  & $-0.314-03$ \\
         &  $X^L_{l}(A)$  & Eq.~(\ref{eq:jta2})   & N2LO & & $+0.875-03$ & & & $+0.687-03$ & \\
         &  $X^E_{l}(A)$  & Eq.~(\ref{eq:jta1})   & LO   & & $+0.597+00$ & & & $+0.613+00$ & \\
         &  $X^M_{l}(B)$  & Eq.~(\ref{eq:jtb1})   & NLO  & & $+0.157-02$ & & & $+0.161-02$ & \\
         &  $X^E_{l}(A)$  & Eq.~(\ref{eq:jta2})   & N2LO & & $-0.124-03$ & & & $-0.978-03$ & \\         
        \hline
      \end{tabular}    
    \end{center}
  \end{table*}

As it can be seen by inspecting the Table, the LO transition operators give the largest RMEs. The dependence of these RMEs from
the nuclear interaction is rather weak. The RMEs coming from NLO and N2LO operators are noticeably suppressed, although
their dependence on the nuclear interaction is more sizable. In any case, in the S, V, A, and T cases, the cumulative RMEs are dominated
by the LO contributions, and therefore, almost no dependence on the interaction is observed. In the P case, the only contribution comes from a N2LO operator, but the dependence on the NN interaction is still weak. 
For the scalar case, we have also calculated the RME coming from the operator given in Eq.~(\ref{J1scalar3}), which, for the AV18 interaction, turns out to be $1.922\times 10^{-2}$, much smaller than the LO, NLO, and N2LO values reported in Table~\ref{tab:rmed}. Therefore, also numerically, we have the confirmation that the contribution of this operator can be safely neglected.

\begin{figure} \centering
    \includegraphics[width=\columnwidth]{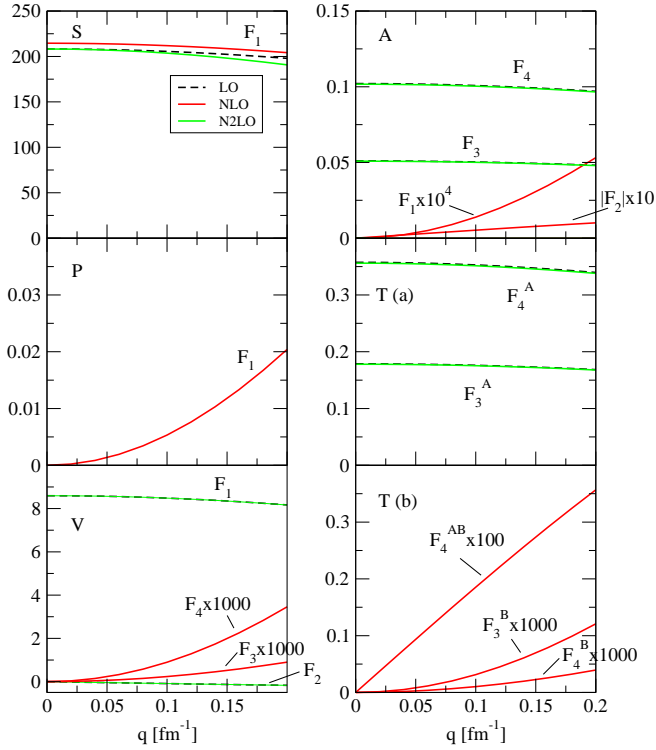}
    \caption{(color online) The cumulative deuteron form factors for the various cases calculated with the
      AV18 potential~\protect\cite{AV18}. Black dashed, solid red, and solid green denotes the form factors calculated
    at LO, NLO, and N2LO, respectively. }
    \label{fig:ff_av18}
\end{figure}

In Fig.~\ref{fig:ff_av18}, the various deuteron form factors calculated with the AV18 potential and the transition currents
at LO, NLO, and N2LO are shown. The form factors are calculated for $q$ values up to $q=0.2$ fm${}^{-1}$
(corresponding to deuterons recoiling with an energy of $q^2/2M_2\approx 390$ keV).
As it can be seen from this figure, the effect of the NLO and N2LO components in the
transition operators are rather tiny, confirming what was shown in Table~\ref{tab:rmed} for the RMEs at $q=0.05$ fm${}^{-1}$.
In the V case, the dominant form factor is $F_1$, while in the A case, the dominant
ones are $F_3$ and $F_4$. For the T case, the dominant form factors are $F_3^A$ and $F_4^A$.

\begin{figure} \centering
    \includegraphics[width=\columnwidth]{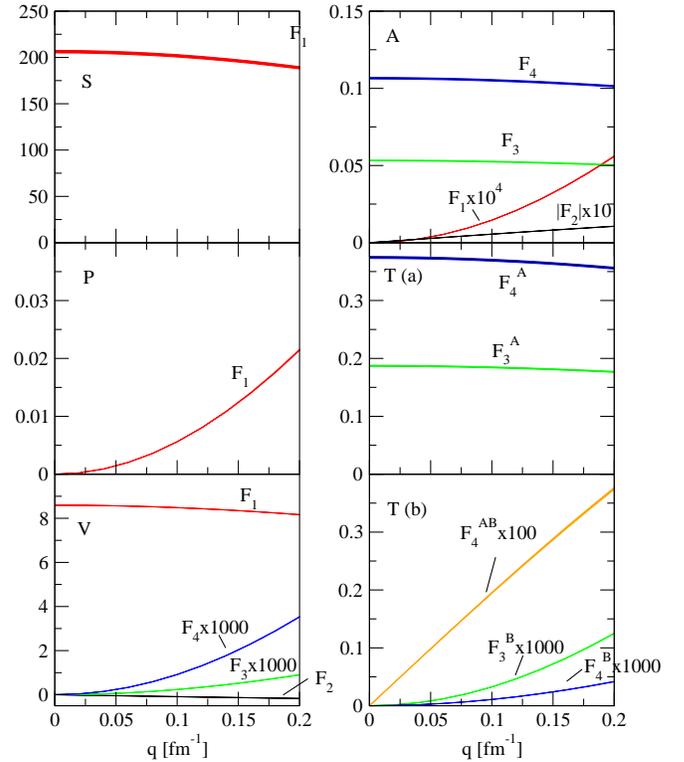}
    \caption{(color online) The deuteron form factors for the various cases calculated with the
      N4LO chiral potentials~\protect\cite{MEN17}. The results are presented as bands, all the calculations are performed
      including the transition operators up to N2LO.}
    \label{fig:ff_chiral}
\end{figure}

In Fig.~\ref{fig:ff_chiral} we report the same form factors calculated with the N4LO450, N4LO500, and N4LO550
NN interactions~\cite{MEN17}. The N4LO450 and N4LO550 are NN potentials derived at N4LO in the framework of $\chi$EFT, but regularized at large momenta with cutoff $450$ and $550$ MeV, respectively.  In all cases, we have used the full transition currents up to N2LO
operators. All the results are shown as band (some of them very narrow), their width reflecting
the spread of theoretical results using the three different cutoff values.
Therefore, the band width reflects our incomplete knowledge of the nuclear dynamics and 
gives a first estimate of the associated theoretical uncertainty. Strictly speaking, such procedure yields only a
lower bound on the theoretical uncertainty~\cite{Furn15a}. In future, we plan to perform a better estimate of such a 
theoretical uncertainty, in particular, using the calculations performed with the interactions
and transition currents at various chiral orders and using the Bayesian procedure of Refs.~\cite{Furn15b,Weso16,Weso21}.
At present we limit ourselves to note that the band widths are rather narrow, so this theoretical uncertainty
seems to be well under control in this kinematic regime of low $q$ values.

\begin{figure} \centering
    \includegraphics[width=\columnwidth]{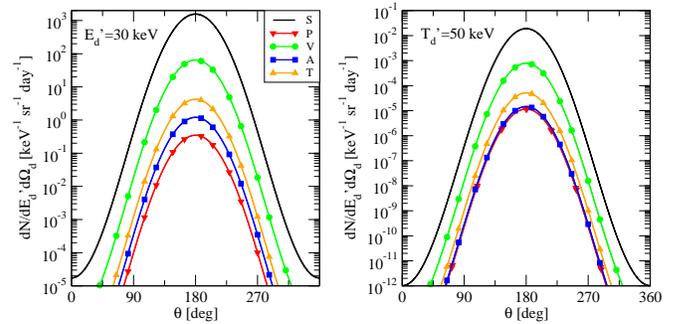}
    \caption{(color online) The number of scattered deuterons per day as function of the angle between $\bmV$ and $\bmP_d'$
      by a 100 ton of deuterium, per unit of energy (keV) and solid angle (sr). For all cases we have taken $M_\chi=10$ GeV and $C^X_+=10^{-4}$
      for the sake of comparison. The left (right) panel reports the number of events for scattered deuterons of recoil energy 30 (50) keV. 
      All the results are presented as (very narrow) bands (see the main text for more detail).}
    \label{fig:rate_chiral}
\end{figure}

In Fig.~\ref{fig:rate_chiral} we report the numbers of scattered deuterons per day as function of the angle between $\bmV$ and $\bmP_d'$
by a 100 ton of deuterium, per unit of energy (keV) and solid angle (sr). This quantity is calculated from Eq.~(\ref{eq:d3rate})
multiplying by the number of second in a day. For all cases we have taken $M_\chi=10$ GeV and $C^X_+=10^{-4}$
for the sake of comparison. Moreover, we have assumed $N_A\approx 3\times10^{31}$ and $N_\chi=6\times 10^{-2}$ cm${}^{-3}$
(calculated from the estimate local energy density of DM $\rho_\chi=0.3$ GeV cm${}^{-3}$\cite{Cooley}). Several observations are in order: 1) there is a large dependence on
the angle $\theta=\cos^{-1}(\hat\bmV\cdot\hat\bmP_d')$, coming from the quantity $A$ in the exponential in Eq.~(\ref{I}).
The rates are peaked at $\theta=180$ deg since a terrestrial target moves with average velocity $\bmV$ in the
(supposed) WIMP cloud: in the laboratory most of the scattered deuterons would be observed to recoil in the
direction $-\bmV$. 2) The number of events also depends critically on the kinetic energy of the detected recoiling  deuteron, the smaller is the better.
3) It also depends on the DM-quark interaction type, assuming the same coupling constant $C^X_+$, $X=S,P,V,A,T$; the largest number of events
would correspond to a scalar coupling between WIMP and quarks; such a type of interactions are already severely constrained
by the existing limits provided by the experiments. 4) The results shown in the figure are actually bands;
the bands gather the rates calculated with the N4LO450, N4LO500, and N4LO550 NN interactions and,
for each interaction, those obtained with transition currents from LO to N2LO; therefore each band includes the
results of nine different calculations (for the P case, three calculations); due to the figure scale, the width of the bands cannot
be well appreciated and this is a confirmation that the results weakly depend on the NN interaction and that the LO transition operators give the dominant contribution. 5) We note that the number
of events for the P case increases relatively to the other cases at $E_d'=50$ keV. This is due to the fact that
the RME for the P case increases as $q^2$, while for the other cases the dominant RMEs are only weakly dependent on $q$.

\begin{figure} \centering
    \includegraphics[width=\columnwidth]{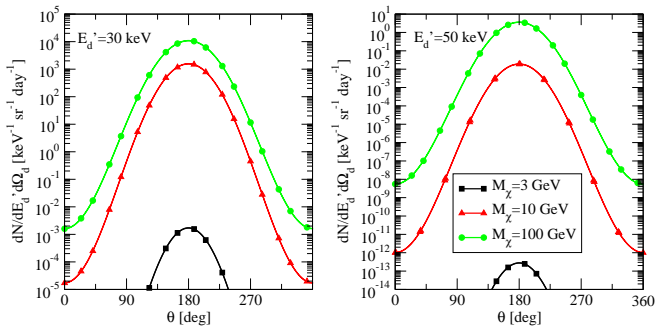}
    \caption{(color online) The number of scattered deuterons per day as function of the angle between $\bmV$ and $\bmP_d'$
      by a 100 ton of deuterium, per unit of energy (keV) and solid angle (sr), for three different values of the WIMP mass $M_\chi$.
      Here we have considered the S interaction with $C^S_+=10^{-4}$.
      The left (right) panel reports the number of events for scattered deuterons of recoil energy 30 (50) keV. 
      All the results are presented as (very narrow) bands (see the main text for more detail).}
    \label{fig:rate_chiral_amx}
\end{figure}

Finally, in Fig.~\ref{fig:rate_chiral_amx} we report the number of events for the S-type interaction for three
different values of the WIMP mass $M_\chi$. Clearly, for lighter WIMPs, the recoil deuterons at a given energy decreases
noticeably. This dependence on the WIMP mass is particularly critical for light WIMP, with mass around 1 to 10 GeV. For mass
greater than 10 GeV the dependence is less relevant. 

\subsection{$\heq$-DM scattering} 
\label{sec{he}}
In the case of the scattering off the $\heq$ nucleus, which has spin $0$, only the multipoles with $l=0$ contribute. Disregarding the very tiny components with negative parity of its wave function, we are left with $X^C_0$ and $X^L_0$ RMEs for the S, V, and T cases. Furthermore, we disregard the RMEs of the isovector operators, since the $\heq$ wave function is with a very good approximation of an almost pure state of total isospin $T=0$. In Table~\ref{tab:rmes4} we report
the calculated RMEs at $q=0.05$ fm$^{-1}$.  The $\heq$ wave functions have been obtained using two interactions: the first is given by the AV18 NN potential augmented by the Urbana IX (UIX) three-nucleon (3N) interaction~\cite{Pudliner97};
the second is given by the N4LO500 NN potential augmented by a N2LO 3N interaction, derived in the framework of $\chi$EFT~\cite{Eea02}.
The two free parameters in this N2LO 3N potential, usually denoted as  $c_D$ and $c_E$, have been
fixed in order to reproduce the experimental values of the $A=3$ binding energies
and the Gamow-Teller matrix element (GTME) of the tritium $\beta$ decay~\cite{GP06,GQN09}.
These parameters have been determined in Ref.~\cite{Mea18}. The cutoff in this 3N interaction has been chosen to be consistent with the corresponding value of the NN interaction, therefore the full NN interaction will be denoted as N4LO/N2LO500. With both interactions, AV18/UIX and N4LO/N2LO500, the experimental $\heq$  binding energy is well reproduced. 

\begin{table}[bth]
  \caption{\label{tab:rmes4}
   The RMEs $X^C_{0}$ and $X^L_{0}$ contributing
   to DM scattering off $\heq$ calculated for AV18/UIX and N4LO/N2LO500 interactions (see the main text for details). Here
   $q=0.05$ fm${}^{-1}$. In the fourth column, we report the order of the transition operator.
   Since the $\heq$ has predominantly zero isospin, only the RMEs of the
   isoscalar operators are reported. For the S and V (T) cases, $X^C$ and $X^L$ are predominantly real (imaginary), and therefore we have reported only those parts. 
   As in Table~\protect\ref{tab:rmed}, the equation numbers reported in the third column 
   specify the operators from which these RMEs are calculated, as
   reported in detail in Appendix~\protect\ref{app:WIMP-N_Int}.}
    \begin{center}
      \begin{tabular}{llll|c|c}
        \hline\hline
         Int. & RME & Operator & order & AV18/UIX & N4LO/N2LO500 \\
         \hline
         S  & & & & $l=0$ & $l=0$ \\
            &  $X^C_{l}$  & Eq.~(\ref{J1scalar0}) & LO   & $-0.169+02$ & $-0.168+02$ \\
            &                & Eq.~(\ref{J2scalar1}) & NLO  & $-0.273+00$ & $+0.144+00$ \\
            &                & Eq.~(\ref{J1scalar2}) & N2LO & $+0.521+00$ & $+0.313+00$  \\
    \hline
         V  & & & & $l=0$  & $l=0$  \\
         &  $X^C_{l}$  & Eq.~(\ref{eq:rho1v1}) & LO   & $-0.343+01$ & $-0.341+01$  \\
         &                & Eq.~(\ref{eq:rho1v3}) & N2LO & $+0.325-04$ & $+0.336-04$ \\
         &  $X^L_{l}$  & Eq.~(\ref{eq:j1v1})   & NLO  & $-0.444-02$ & $-0.438-02$ \\
        \hline
         T  & & & & $l=0$  & $l=0$  \\
         &  $X^L_{l}(B)$  & Eq.~(\ref{eq:jtb1})   & NLO  & $-0.315-02$ & $-0.325-02$ \\
        \hline
      \end{tabular}    
    \end{center}
  \end{table}

\begin{figure} \centering
    \includegraphics[scale=0.5]{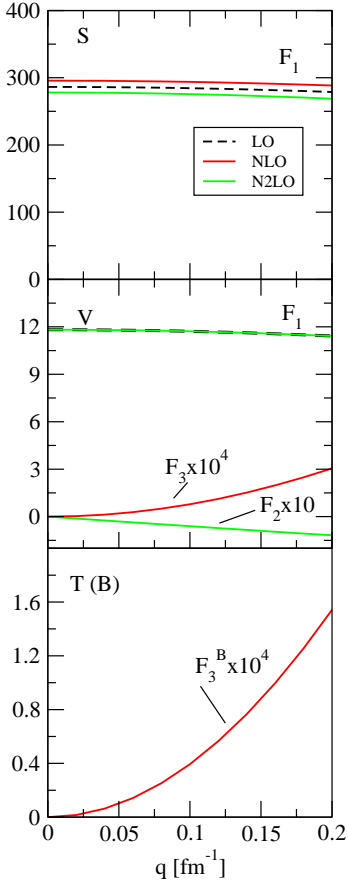}
    \caption{(color online) The same as Fig.~\ref{fig:ff_av18} but for $\heq$.}
    \label{fig:ff_4he_av18}
\end{figure}

In Fig.~\ref{fig:ff_4he_av18}, the various $\heq$ form factors calculated  for $q$ values up to $q=0.2$ fm${}^{-1}$ with the AV18 potential and the transition currents at LO, NLO, and N2LO are shown. For $q=0.2$ fm${}^{-1}$, the $\heq$ recoil energies $q^2/2M_4\approx 195$ keV.
As it can be seen from this figure, for the S case, the effect of the NLO and N2LO components in the
transition operators are more sizeable, while in the V case rather tiny. The only contribution for the T case now comes from a NLO transition current, therefore the only non-vanishing form factor, $F^{3,B}(q)$, is very small and varying as $q^2$. We expect therefore that the rate for the T case to depend noticeably on the $\heq$ recoil energy.  

\begin{figure} \centering
    \includegraphics[scale=0.5]{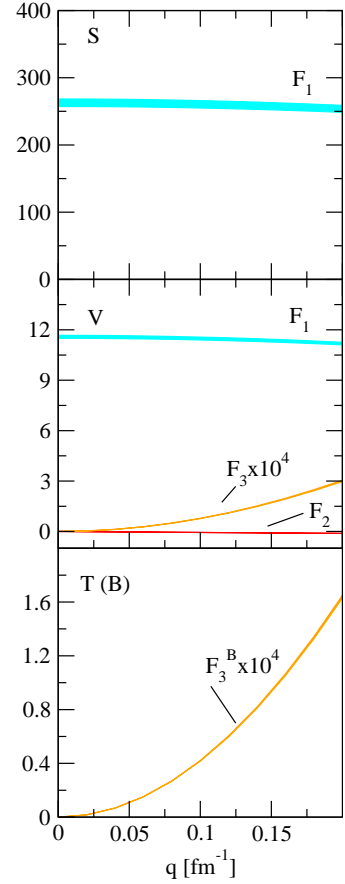}
    \caption{(color online) The same as Fig.~\ref{fig:ff_chiral} but for $\heq$.}
    \label{fig:ff_4he_chiral}
\end{figure}

In Fig.~\ref{fig:ff_4he_chiral} we report the same form factors calculated with the N4LO/N2LO450, N4LO/N2LO500, and N4LO/N2LO550 NN interactions (as specified, for each N4LO NN interaction we have added the N2LO 3N interaction regularized with the same cutoff). As for the deuteron case, we have used the transition currents up to N2LO and the results are shown as band (some of them very narrow), their width reflecting
the spread of theoretical results using $\Lambda=450$, $500$, or $550$ MeV cutoff values.
The width for the S case is sizeable, while for all other cases they are practically negligible. 

\begin{figure*} \centering
    \includegraphics[scale=.5]{rate_4he_chiral.eps}
    \caption{(color online) The same as Fig.~\ref{fig:rate_chiral} but for $\heq$. The horizontal blue lines denote the number of scattered $\heq$ per day due to the background neutrino flux of atmospheric origin, see Subsect.~\protect\ref{sec:nu} for more details. }
    \label{fig:rate_4he_chiral}
\end{figure*}

\begin{figure*} \centering
    \includegraphics[scale=.5]{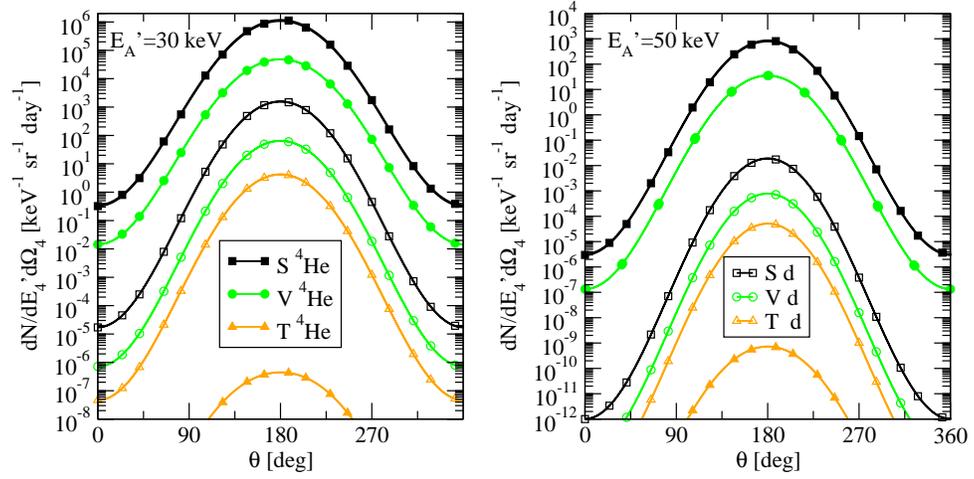}
    \caption{(color online) Comparison between the rates for deuteron and $\heq$.
    For all cases we have taken $M_\chi=10$ GeV and $C^X_+=10^{-4}$. Full symbols: $\heq$ rates; open symbols: deuteron rates.}  \label{fig:rate_d_4he_chiral}
\end{figure*}

In Fig.~\ref{fig:rate_4he_chiral} we report the numbers of scattered $\heq$ per day as function of the angle between $\bmV$ and $\bmP_d'$ by a 100 ton of $\heq$, per unit of energy (keV) and solid angle (sr). For all cases we have taken $M_\chi=10$ GeV and $C^X_+=10^{-4}$
for the sake of comparison. Now, $N_A\approx 3\times10^{31}$, while we have kept $N_\chi=6\times 10^{-2}$ cm${}^{-3}$. A similar behavior is observed as in Fig.~\ref{fig:rate_chiral},
with the only difference that now the rate for the T case is suppressed, due to the previously discussed very small size of the (only contributing) form factor $F^{3,B}(q)$. In this figure, we have also reported the number of scattered $\heq$ per day due to the background neutrino flux of atmospheric origin, see Subsect.~\protect\ref{sec:nu} for more details.
Furthermore, a comparison between the rates for deuteron and $\heq$ is shown in Fig~\ref{fig:rate_d_4he_chiral}.

\subsection{Rate for $\nu-\heq$ scattering}
\label{sec:nu}

An important  background for DM experiments is that given by the nuclear recoils due to the flux of neutrinos. These neutrinos may have different origins. For nuclear recoils in the range $30$-$50$ keV the most important flux is due to atmospheric neutrinos~\cite{nu}. The differential cross-section for $\nu-\heq$ can be calculated starting from the following effective Lagrangians
\begin{equation}
    {\cal L}= - {G_V\over \sqrt{2}} j_\mu^\ell J^\mu_N\ ,
    \label{eq:nu1}
\end{equation}
where $G_V$ is the Fermi constant, $j_\mu^\ell=\overline\psi_\nu \gamma_\mu(1-\gamma^5)\psi_\nu$ the leptonic neutrino current, and $J^\mu_N$ the neutral nuclear current.
The latter quantity can be obtained using the $\chi$EFT approach, following the same lines detailed in this paper. In the present paper, we consider the neutral weak current derived at N4LO in Refs~\cite{baroni15,Pastore09,Shen12}. The differential $\nu-\heq$ cross-section is given by
\begin{eqnarray}
    {d^2\sigma\over dE_4' d \Omega_4} &= & {G^2_V\over\pi} P_4' M_4 \Bigl[
    (1+\hat k\cdot\hat k') |C_0|^2 \nonumber\\
    &&+(1-\hat k\cdot\hat k'+2 \hat k\cdot\hat q \hat k'\cdot \hat q) |L_0|^2\nonumber\\
    && - (\hat k\cdot \hat q+\hat k'\cdot\hat q)2 {\text{Re}}(C_0 L_0^*)
    \Bigr]\nonumber\\
    &&\times \delta(k-E_4'-k')\ ,
\end{eqnarray}
where $\bmk$ ($\bmk'$) is the incoming (outgoing) neutrino momentum, while
$E_4'$ and $\bmP_4'=\bmk-\bmk'\equiv \bmq$ the kinetic energy and momentum of the recoiling $\heq$ nucleus. Moreover, $C_0$ and $L_0$ are the RMEs calculated from the matrix elements $\langle \Psi_4|J^\mu_N|\Psi_4\rangle$. The factors multiplying the combinations of RMEs comes from the traces of the lepton currents over the neutrino spins. The energy conservation imposes that $\bmk\cdot\bmP_4'=T(k+M_4)$, where hereafter $T=E_4'$. Typically the neutrino energies are in the range of MeV, while $T\approx$ keV. So disregarding all terms proportional to $T/k$ and integrating over $\Omega_4'$ one obtains the typical cross-section for neutrino-nucleus scattering~\cite{nu}
\begin{equation}
    {d\sigma\over dT}= {G_V^2\over 4\pi} M_4 Q^2_W \left(1-{TM_4\over 2k^2}\right) F_W(q)\ ,
\end{equation}
where $F_W(q)=4\pi |C_0|^2/(Q^2_W/4)$ and $Q_W=2(1-4\sin^2\theta_W)-2$,
being the $\heq$ "weak charge"  (in this way $F_W(0)\approx 1$). 
Above $\theta_W$ is the Weinberg angle, $\sin^2\theta_W\approx 0.223$.

The flux of atmospheric neutrinos is nearly isotropic, peaks at $k=k_0\approx 30$ MeV, and at $k=10^3$ MeV is reduced by a factor $100$~\cite{nu,Newstead21}. Then, the rate of $\heq$ recoils due to the flux of atmospheric neutrinos is given by
\begin{equation}
    {d^2R\over dT d\Omega_A'}= N_A \int {d\hat k\over 4\pi}\, dk \,{d\phi(k)\over dk} {d^2\sigma\over dT d\Omega_4'}\ ,
\end{equation}
where $N_4$ is the number of $\heq$ nuclei in the target (we assume as before to have a 100 ton target). Moreover,  we approximate $d\phi(k)/dk\approx \phi_0 \exp[-(k-k_0)^2/(2\sigma_\nu)^2]$, where $\phi_0\approx 10^{-2}$ cm${}^{-2}\,$ MeV${}^{-1}\,$ sec${}^{-1}$~\cite{nu}, and
$\sigma_\nu=226$ MeV, so that $d\phi(10^3\, \text{MeV})/dk=10^{-2}\phi_0$.
Note that the uncertainty on this
atmospheric neutrino flux is approximately 20\% \cite{nu}, therefore the rates calculated in the following have to be considered as order-of-magnitude estimates.

The expected number of events due to the atmospheric neutrinos in a day calculated for $T=30$ keV and $50$ keV have been reported in Fig.~\ref{fig:rate_4he_chiral}. This number is clearly isotropic with respect to the direction of $\bmV$ and it is of order of $10^{-11}$ events per keV and per steradiant. From this number, we can estimate the minimal values for the Wilson coefficients $C^X_+$ which can be measured in an experiment, asking that $d^2R_{DM}$ at $\theta=180$ deg be greater than $d^2R_{\nu}$. The obtained results are reported in Table~\ref{tab:Cnu}. 

\begin{table}[bth]
  \caption{\label{tab:Cnu}
   Minimal values of the Wilson coefficients $C^X_+$ so that the rate
   of recoiling $\heq$ nuclei with kinetic energy $T$ due to DM be greater than that due to the neutrino scattering. The calculations have been performed using the N4LO/N2LO500 potential. Here we have considered $M_\chi=10$ GeV. 
  }
    \begin{center}
      \begin{tabular}{l|ccc}
        \hline\hline
         $T$ (keV) & $C^S_+$ & $C^V_+$ & $C^T_+$ \\
         \hline
         $30$  & $4.5\times 10^{-13}$ & $2.1\times10^{-12}$ & $7.0\times 10^{-7}$ \\
         $50$  & $1.6\times 10^{-7}$ & $7.7\times 10^{-7}$ & $1.7\times10^{-1}$ \\
         \hline
      \end{tabular}    
    \end{center}
  \end{table}

\section{Conclusions}
\label{sec:conc}

In this paper, we have studied the scattering of WIMP off some light nuclei. The aim is twofold. First of all, we have explicitly written down most of the transition currents for various types of DM-quark interactions, assuming DM is composed of heavy Dirac particles. The transition currents, developed up to N2LO in the framework of $\chi$EFT,  have been coded in a way to be used for a general nuclear system, as, for example, using the shell model approach. Second, we have set up the calculation directly for the rate of nuclear recoils, in order to be ready for a direct comparison with (eventual) experimental yields. We have also set up the calculation of the rate induced by the flux of terrestrial or cosmological neutrinos (in particular, atmospheric), calculating the matrix elements of the nuclear neutral current. 

We have performed calculations for two targets composed either of deuterons or $\heq$ nuclei, the latter nucleus being actually considered  for an experiment~\cite{heliumdm}. The deuteron has total isospin $T=0$, and also the ground state of $\heq$ can be well approximated to have total isospin $T=0$. Therefore, only the isoscalar transition currents play a role in these cases. We have found that the scalar and vector interactions give large values for the form factor $F_1$,  deriving from the matrix elements of the operator $\sum_{j=1}^A e^{i\bmq\cdot\tilde\bmr_j}$ (multiplied by some combinations of LECs). These matrix elements for small values of $q$ are therefore proportional to the number of nucleons $A$; they are not difficult to be calculated also for large nuclei. Clearly, in these cases, the rates would be rather large unless the corresponding Wilson coefficients $C^S_+$,  $C^V_+$ be extremely small. For other interactions, the rate is found to be suppressed (in particular, for a purely pseudoscalar quark-DM interaction). In those cases, the form factors derive from the matrix elements of more complicated operators and therefore sophisticated nuclear structure calculations would be necessary.

Regarding the construction of the nuclear wave functions,  we have limited ourselves to employ the AV18 potential and some chiral interactions differing for their cutoff value. In this way we have explored the dependence of the results on the nuclear interaction, giving a first idea of the theoretical uncertainty related to our not complete knowledge of this quantity. In future, we plan to perform more detailed study using the Bayesian formalism~\cite{Weso21}. Regarding the convergence of the chiral expansion of the transition current, this appears to be well under control, due to the low $q$ values involved in the processes.  

For the scalar case, for both the deuteron and $\heq$ targets, we find that the NLO two-body currents modify the LO results by a few percent only, as found in Ref.~\cite{Vries23} and that their contribution is rather dependent on the nuclear interaction used to calculate the ground-state wave functions. A similar result is again observed in Ref.~\cite{Vries23}, where the effect was traced back to the dependence on the $D$-wave percentage of the wave functions.

In perspective, we plan to apply this formalism to study the rate of DM scattering off heavier nuclei like Argon and Xenon, currently widely used in DM detectors. We plan also to study other possible types of DM interactions, as direct couplings to photons (as, for example,  ${\cal L}\approx \overline\chi\sigma_{\mu\nu}\chi F^{\mu\nu}$, $F^{\mu\nu}$ being the electromagnetic field)~\cite{PanciPC}. The extension of the present formalism to treat either scalar or Majorana WIMPs is also possible~\cite{Goodman_2011,Bishara17}. 

Finally, in the last years, the idea of light DM has gained more credit (see, for example, Ref.~\cite{FIP23}). The only change in our formalism is the calculation of the spin sums given in Eq.~(\ref{LL}), which for light DM can be performed directly using the trace formalism, as for the neutrino case. Therefore, the present study could be easily extended to treat such a case. 

\begin{acknowledgements}
The Authors would like to acknowledge L. Girlanda (University of Salento) and P. Panci (Universiy of Pisa)
for useful discussions.
\end{acknowledgements}

\appendix
\section{WIMP-Nucleon Interactions}\label{app:WIMP-N_Int}
\subsection{Scalar interaction}

The scalar interaction is characterized by an external current $s(x)$. 
The terms of the chiral Lagrangian containing this quantity are~\cite{fettes00}
\begin{eqnarray}\label{lagrc1c5}
\mathcal{L}^S_{int}&=&\mathit{c}_1\bar{N}\langle \xi_+\rangle N+\mathit{c}_5\bar{N}\hat{\xi}_+ N\nonumber\\&+&\frac{f_\pi^2}{4}\left\langle \xi(x) U^\dagger(x)+ U(x) \xi^\dagger(x)\right\rangle+\cdots
 \end{eqnarray}
where  $c_1$ and $c_5$ are LECs and
\begin{eqnarray}
\xi_+ &=& u^\dag \xi \,u^\dag + u\, \xi^\dag u\,,\\
\xi(x)&=&2B_c\,s(x)\,,\\
U(x)&=&e^{\frac{i}{f_\pi}\vec\pi(x)\cdot \vec\tau}\,,\\
u(x)&=& \sqrt{U(x)}\,.
\end{eqnarray}
Above $B_c$ is another LEC related to the pion mass value. 
In Eq.~(\ref{lagrc1c5}) $\left\langle \dots \right\rangle$ indicates the trace of the matrices, $\hat{A}=A-\frac{1}{2}\langle A \rangle$
and the dots  represent higher order terms which are negligible for our purposes.
Explicitly, the WIMP contribution to the density $s(x)$ is given by Eq~(\ref{scurr}).
Expanding the Lagrangian~(\ref{lagrc1c5}) in power of the pion field and keeping the terms up to the order $O(\pi^2)$,
as it will result clear later, we need only to consider the following Lagrangian terms
\begin{alignat}{2}
  \mathcal{L}_{int}=-& 8B_c c_1 \frac{C^S_+}{\Lambda_S^2}\bar{N}N\bar{\chi}\chi-
  4B_cc_5 \frac{C^S_-}{\Lambda_S^2}\bar{N}\tau_z N \bar{\chi}\chi\nonumber\\
  +&B_c \frac{C^S_+}{\Lambda_S^2}\bar{\chi}\chi \pi^2\,.\label{finalscalarlag2}
\end{alignat}
The interaction Hamiltonian can be obtained from the chiral Lagrangian density  using the procedure
described in detail in Ref.~\cite{baroni15}. In most of the cases,  the Hamiltonian terms are simply given by, 
\begin{equation}
\mathcal{H}_{int}(x)=-\mathcal{L}_{int}(x),
\end{equation}
but in special cases there are correction terms to be taken into account.

The WIMP current in this case is given by
\begin{equation}
  L_{\bmk' r',\bmk r}^\mu = \bar u^{\chi}_{\bmk' r'} u^{\chi}_{\bmk r}\,,\label{eq:LS}
\end{equation}
where $u^\chi_{\bmk r}$ are Dirac four-spinors. Expanding these latter quantities in powers of the momenta, we have
\begin{equation}
  L_{\bmk' r',\bmk r}=\left( 1-\frac{(\bmk + \bmk')^2}{8M_\chi^2}\right)\delta_{r,r'}-\frac{i(\bmk'\times \bmk)\cdot\bmsi_{r'r}}{4M_\chi^2}
  + \cdots\,,
\end{equation}
where  $\bmsi_{r'r}$ denotes the matrix element of the Pauli matrices between the WIMP spin states.

\begin{figure}[h!]
  \centering
  \includegraphics[scale=0.8]{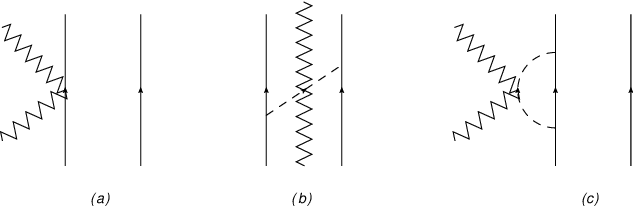}
  \caption{Diagrams contributing to the one-- and two-body transition operators for the scalar case.
    Solid, dashed, and wiggly lines represents nucleons, pions, and WIMPs, respectively. Only one
    time-ordering of each diagram has been reported.}
  \label{fig:scalar}
\end{figure}

In Fig.~\ref{fig:scalar}, the diagrams contributing to the one-- and two-body transition operators for the scalar case
have been reported. The diagram depicted in panel (a) gives a LO contribution of order $Q^{-3}$ plus an additional
N2LO contribution of order $Q^{-1}$ coming from the expansion of the Dirac four-spinors entering the $NNWW$ vertex.
The other two diagrams contribute to NLO. Corrections to these two diagrams due to the expansion of the Dirac four-spinors and
the energy denominators are at least of order $Q^0$ and therefore we will neglect them. Considering all
contributions up to N2LO, the one- and two-body and the WIMP currents are given by
\begin{alignat}{2}
  &J_{\alpha,\alpha'}^{(1)}=\left(\frac{8B_cc_1C^S_+}{\Lambda_S^2}\delta_{t't}+\frac{4B_cc_5C^S_-}{\Lambda_S^2}(\tau_z)_{t't}\right)\nonumber\\
  &\hspace{1cm}\cdot\bigg[\left( 1-\frac{(\bmp + \bmp')^2}{8M^2}\right)\delta_{s,s'}-\frac{i(\bmp'\times \bmp)\cdot\bmsi_{s's}}{4M^2}\bigg]
  \nonumber\\
  &\hspace{1cm}+\frac{C^S_+}{\Lambda_S^2}\frac{3g_A^2B_cm_\pi}{32\pi f_\pi^2}F(\frac{q}{2m_\pi})\delta_{t't}\delta_{s's}\,,\label{J1scalar}\\
  &J_{\alpha_1\alpha_2,\alpha_1',\alpha_2'}^{(2)}=
  -\frac{C^S_+}{\Lambda_S^2}B_c\frac{g_A^2 }{(2 f_\pi)^2}
  \bigg(2\vec{\tau}_1\cdot\vec{\tau}_2\,\frac{i\bmk_1\cdot\bmsi_1\, i\bmk_2\cdot\bmsi_2}{\omega_{k_1}^2\omega_{k_2}^2}\bigg)\label{J2scalar}\,,
\end{alignat}
where $\bmsi_{s's}$ ($\bmta_{t't}$) is the matrix element of the Pauli matrix $i=x,y,z$ between nucleon spin (isospin) states. Moreover, hereafter
we also use the notation $\bmsi_1\equiv (\bmsi_{s_1',s_1})$, $\omega_k=\sqrt{m_\pi^2+k^2}$, etc. 
Above $F(x)=\frac{(2x^2+1){\text{arctan}}(x)+2x}{x}$ derives from the dimensional regularization of the pion loop in panel (c) of
Fig.~\ref{fig:scalar}, see Ref.~\cite{Korber17} and references therein.

For deuteron and $\heq$ scattering, only the isoscalar part of the
transition current will play a role, so let us write explicitly its chiral components, order by order. Let us define
\begin{equation}
    J^{(1)}_{\alpha,\alpha',\mathrm{is}}=\frac{C^S_+}{\Lambda_S^2}\left[ \sum_{\nu=0,2} J^{(1),\nu}_{\alpha,\alpha',\mathrm{is}}\right] \,, 
\end{equation}
where the subscript ''is'' specifies that we are considering the isoscalar part, only. It is also convenient to introduce the so-called ``$\sigma$'' term~\cite{Korber17}, defined as
$\sigma_{\pi N}=\langle N|\bar q q |N\rangle$. In $\chi$EFT, $\sigma_{\pi N}$ is given by~\cite{Korber17}
\begin{equation}
  {\sigma_{\pi N}\over m_\pi} = 4 c_1 m_\pi + \frac{9\pi g_A^2 m_\pi^2}{4 (4\pi f_\pi)^2} A(\frac{q}{2m_\pi})+\cdots
  \,, \label{sigmapin}
\end{equation}
where the ``$\cdots$'' denote high order terms in $\chi$PT and  $A(x) ={1\over3}F(x)-1$. The LO, NLO, and N2LO one-body isoscalar components can be re-cast in the form~\cite{Korber17}
\begin{eqnarray}
   J^{(1),0}_{\alpha,\alpha',\mathrm{is}} &=&   {2\sigma_{\pi N} B_c\over m_\pi^2} \delta_{s,s'} \delta_{t,t'} \,,\label{J1scalar0} \\
  J^{(1),1}_{\alpha,\alpha',\mathrm{is}} &=& 0\,,\label{J1scalar1} \\
  J^{(1),2}_{\alpha,\alpha',\mathrm{is}} &=&    {2\sigma_{\pi N} B_c\over m_\pi^2}  \Big[ -\frac{(\bmp + \bmp')^2}{8M^2} \delta_{s,s'}\nonumber\\
   &&-\frac{i(\bmp'\times \bmp)\cdot\bmsi_{s's}}{4M^2}\Big]  \delta_{t,t'} \,.\label{J1scalar2} 
\end{eqnarray}
The NLO term coming from the last term of Eq.~(\ref{J1scalar}) has been absorbed in the definition of $\sigma_{\pi N}$, while the rest, proportional to $A(x)$, gives a term which reads
\begin{equation}
 J^{(1),3}_{\alpha,\alpha',\mathrm{is}} =  \frac{9g_A^2B_cm_\pi}{64\pi f_\pi^2}A(\frac{q}{2m_\pi}) \delta_{s,s'} \delta_{t,t'}\,.\label{J1scalar3}
\end{equation}
However, this term is  of third order, as $A(x)\approx x^2$, and therefore it will be neglected in this work (see also later). 
The isoscalar two-body current can be decomposed in the same way
\begin{eqnarray}
  J^{(2)}_{\alpha_1\alpha_2,\alpha_1',\alpha_2',\mathrm{is}}&=& \frac{C^S_+}{\Lambda_S^2} \sum_{\nu=0,2} J_{\alpha_1\alpha_2,\alpha_1',\alpha_2',\mathrm{is}}^{(2),\nu}\,, \\
  J^{(2),1}_{\alpha_1\alpha_2,\alpha_1',\alpha_2',\mathrm{is}}&=& - B_c\frac{g_A^2 }{2f_\pi^2}
    \vec{\tau}_1\cdot\vec{\tau}_2\,\frac{i\bmk_1\cdot\bmsi_1\, i\bmk_2\cdot\bmsi_2}{\omega_{k_1}^2\omega_{k_2}^2}\label{J2scalar1}\,,
\end{eqnarray}
and clearly $J^{(2),0}=J^{(2),2}=0$.

\subsection{Pseudoscalar Interaction}\label{pseudoscalarint}
The pseudoscalar external current $p(x)$, given in Eq.~(\ref{pcurr}), enters  the effective nucleonic Lagrangian through the following operators~\cite{fettes00}
\begin{eqnarray}
  \xi &=&2iB_cp(x)\\
  \xi_{\pm}&=& u^\dag \xi \,u^\dag \pm u\, \xi^\dag u\,,\label{chipm}
\end{eqnarray}
The interaction Lagrangian is given explicitly as~\cite{fettes00}:
\begin{alignat}{2}
\mathcal{L}_{int }^{P}&=\frac{f_\pi^2}{4}\langle\xi(x)U^{\dagger}(x)+U(x)\xi^{\dagger}\rangle \nonumber \\&+  d_{18} \bar{N}{i\over 2}
  \gamma^\mu\gamma^5 [\partial_\mu, \xi_-] N\nonumber \\
 &+d_{19} \bar{N}{i\over 2}
  \gamma^\mu\gamma^5 [\partial_\mu, \langle\xi_-\rangle] N+\dots\ .\label{plagr}
\end{alignat}
Expanding the above Lagrangian in powers of the pion field, we obtain
\begin{alignat}{2}
\mathcal{L}_{int }^P &=2f_\pi B_c\frac{C^P_-}{\Lambda_S ^2}\bar{\chi}i\gamma^5\chi\pi_z \nonumber\\&-2B_c\frac{C^P_+}{\Lambda_S ^2}(d_{18}+2d_{19})\bar{N}\gamma^\mu\gamma^5 N\; \partial_\mu (\bar{\chi}i\gamma^5\chi)\nonumber\\
&-2B_c\frac{C^P_-}{\Lambda_S ^2}d_{18}\bar{N}\gamma^\mu\gamma^5 \tau_z N \; \partial_\mu (\bar{\chi}i\gamma^5\chi)\,,
\end{alignat}
where $d_{18}, d_{19}$ are LECs.

The WIMP current in this case is given by
\begin{equation}
  L_{\bmk' r',\bmk r} = \bar u^{\chi}_{\bmk' r'} i\gamma^5 u^{\chi}_{\bmk r}\,,\label{eq:LP}
\end{equation}
and expanding the Dirac four-spinors in powers of the momenta,
\begin{equation}
  L_{\bmk' r',\bmk r}=  i \frac{\bmsi_{r'r}\cdot\bmq}{2M_\chi}+\ldots\,.\label{eq:LP2}
\end{equation}
We note that $ L_{\bmk' r',\bmk r}$ is at least of order $Q$. 

\begin{figure}[h!]
  \centering
  \includegraphics[scale=0.5]{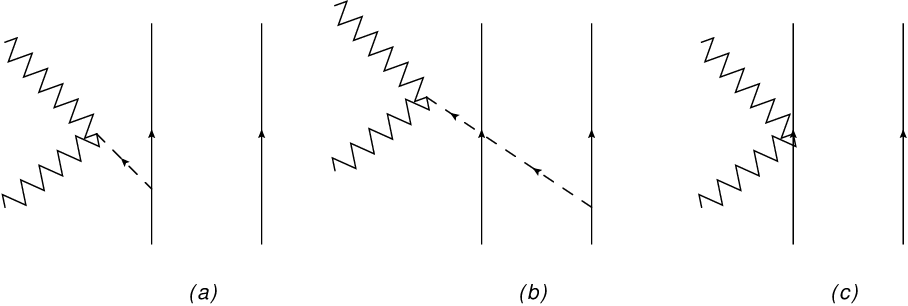}
  \caption{The same as in Fig.~\protect\ref{fig:scalar} but for the pseudoscalar
  interaction. }
  \label{fig:pseudoscalar}
\end{figure}

The relevant diagrams are reported in Fig.~\ref{fig:pseudoscalar}. The LO contribution is given by the
diagram depicted in panel (a) where the WIMP is scattered after absorbing a pion-in-flight. It brings a contribution of order $Q^{-3}$ (hereafter we include in the counting one $Q$ coming from $L_{\bmk' r',\bmk r}$), plus  corrections at order $Q^{-1}$ coming from the expansion of the vertex functions and energy denominators.
However, these latter terms are neglected here since they correspond to isovector transition operators,
which contribution vanishes in the deuteron (for $\heq$ the pseudoscalar coupling does not give any contribution). Here, we keep pseudoscalar isosvector operators at LO only.
The diagram of panel (b) gives again an isovector transition operator of order $Q^{-1}$, therefore, in this work we neglect it, as well. 

The diagram of panel (c) derives from the 
vertex with the $d_{18}$ and $d_{19}$ LECs. It is of order $Q^{-1}$, but it has an isoscalar term, so we take it into account. Other diagrams contribute at order ${\cal O}(Q^0)$ or higher. The final expression of the transition density we will consider is therefore

\begin{alignat}{2}
  \!\!J_{\alpha,\alpha'}^{(1)}&=\!\bigg(\!2B_c\frac{C^P_-}{\Lambda_S ^2}d_{18}(\tau_z)_{t't}\nonumber\\
  &\quad +2B_c(d_{18}+2d_{19})
  \frac{C^P_+}{\Lambda_S ^2}\delta_{t't}\!\bigg) i\bmsi_{s's}\cdot\bmq\nonumber\\
  &+ g_A B_c\frac{C^P_-}{\Lambda_S ^2}\frac{(\tau_z)_{t't}}{\omega_q^2} i\bmsi_{s's}\cdot\bmq
   \,,\label{JeLpseudo}
\end{alignat}
while, as explained above, we neglect the pseudoscalar two-body current in this study.

The isoscalar part, relevant in this study, is rewritten as
\begin{eqnarray}
  J_{\alpha,\alpha',is}^{(1)}&=& {C^P_+\over \Lambda_S ^2} \sum_{\nu=0,2} J_{\alpha,\alpha',is}^{(1),\nu}\,,\\
  J_{\alpha,\alpha',is}^{(1),2}&=& 2B_c(d_{18}+2d_{19}) i\bmsi_{s's}\cdot\bmq\; \delta_{t't} \,,\label{JeLpseudo1}
  \end{eqnarray}
  while $J^{(1),0}=J^{(1),1}=0$.

\subsection{Vector Interaction}\label{vectinteraction}
In the case of vector interaction, the WIMP field contributes to the quantities $v^{(s)}_\mu$ and $v_\mu$,
entering the following Lagrangian terms~\cite{fettes00} 
\begin{eqnarray}
  \mathcal{L}_{int}^V&=&i\bar{N}\gamma^\mu( \Gamma_\mu-i v^{(s)}_\mu)N \nonumber\\
  &+&\frac{\mathit{c}_6}{8M}\bar{N} \sigma^{\mu\nu} F_{\mu\nu}^{+}N+\frac{\mathit{c}_7}{4M}\bar{N} \sigma^{\mu\nu}
  F_{\mu\nu}^{(s)} N \nonumber \\
  &+&\frac{f_\pi^2}{2} \langle \partial_\mu U^\dag ( i U v_\mu-i v_\mu U)\rangle +\cdots \label{eq:LintV2}
\end{eqnarray}
with
\begin{eqnarray}
  \Gamma_\mu &=& {1\over 2} \big[ u^\dag \partial_\mu u + u \partial_\mu u^\dag -i u^\dag v_\mu u - i u v_\mu u^\dag\big]\,,\\
  F^{(s)}_{\mu\nu} &=& \partial_\mu v^{(s)}_\nu  - \partial_\nu v^{(s)}_\mu\ , \\
  F_{\mu\nu} &=& \partial_\mu v_\nu  - \partial_\nu v_\mu\ .
\end{eqnarray}
Above we have neglected all terms quadratic in $v_\mu$ because we suppose that the coupling constants $C_{V\pm}$ are very small.
The LECs $c_6$ and $c_7$ are related to the anomalous magnetic moment of the nucleons.
In the present case $J^{(1)}$, $J^{(2)}$, and $L$ are four vectors, the chiral order of their
``time'' and ``space'' parts being different. Explicitly, the WIMP current in this case is given by,
\begin{equation}
  L_{\bmk' r',\bmk r}^\mu = \bar u^{\chi}_{\bmk' r'}\gamma^\mu u^{\chi}_{\bmk r}\ ,\label{eq:LV}
\end{equation}
which can be expanded up to order $Q^2$ as follows
\begin{eqnarray}
(L_{\bmk' r',\bmk r})^{\mu=0}&\!\!=\!\!&\Bigg[1-\frac{q^2}{8M^2_\chi}
\bigg] \delta_{r'r} +\frac{i(\bmk'\times\bmk)\cdot\bmsi_{r'r}}{4M^2_\chi}\,,\label{eq:LV1}\\
(L_{\bmk' r',\bmk r})^{\mu=i}&\!\!=\!\!&\frac{(\bmk'+\bmk)_i}{2M_\chi}\delta_{r'r}+\frac{\big(i\bmq\times\bmsi_{r'r}\big)_i}{2M_\chi}\,,\label{eq:LV2}
\end{eqnarray}
for $i=x,y,z$. 
As it can be seen $(L_{\bmk' r',\bmk r})^{\mu=0}\approx O(Q^0)$, while $(L_{\bmk' r',\bmk r})^{\mu=i}\approx O(Q)$. 

\begin{figure}[h!]
  \centering
  \includegraphics[scale=0.5]{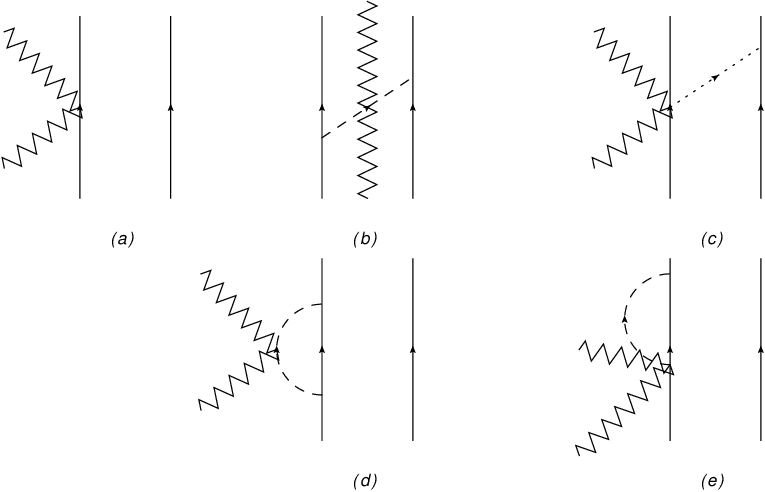}
  \caption{The same as in Fig.~\protect\ref{fig:scalar} but for the vector
  interaction. }
  \label{fig:vector}
\end{figure}

The relevant diagrams for the vector interaction are reported in Fig.~\ref{fig:vector}.
The WIMP-nucleon vertex appearing in the diagram of panel (a)
derives from the interaction terms reported in the first two lines of Eq.~(\ref{eq:LintV}). 
The minimal order of the ``time'' component of $J^{(1)}$
associated to this diagram is $\approx Q^{-3}$, while the space part is $\approx Q^{-2}$. 

The diagrams (b) and (c) give the contribution of the one-pion exchanges. The time part of these diagrams is
always of order $Q^0$. The spatial parts of the corresponding $J^{(1)}$ and $J^{(2)}$ are of order $Q^{-1}$,
so we take into account them. The spatial parts of the diagrams in panels (d) and (e) are of order $Q^{-1}$,
as well. Since they give the first contribution to the form factors of the nucleon, we include them by inserting the ``phenomenological'' electric and magnetic form factors $G_E(q)$ and $G_M(q)$ in our transition currents. In this way,
we take into account also of high order contributions. 
The final expression of the transition densities we will consider is 
\begin{eqnarray}
  (J_{\alpha,\alpha'}^{(1)})^{\mu=0}&=&-h_V(q)\delta_{s's}-\big (2\tilde h_V(q)-h_V(q)\big) \nonumber\\
  && \times \left(-\frac{q^2}{8M^2}\delta_{s's}+\frac{i(\bmp'\times\bmp)\cdot\bmsi_{s's}}{4M^2}\right)\,, \label{eq:rho1v}\\
  (J_{\alpha,\alpha'}^{(1)})^{\mu=i}&=&-h_V(q) \frac{(\bmp+\bmp')_i}{2M}\nonumber\\
  &&+\tilde h_V(q) \frac{i(\bmq\times\bmsi_{s's})_i}{2M}\,,\label{eq:j1v}
\end{eqnarray}
and 
\begin{eqnarray}
  (J_{\alpha_1,\alpha_1',\alpha_2,\alpha_2'}^{(2)})^{\mu=0}&=&0\,,\label{eq:rho2v}\\
  (J_{\alpha_1,\alpha_1',\alpha_2,\alpha_2'}^{(2)})^{\mu=i}&=& {g_A\over 2 f_\pi^2} {C^V_-\over \Lambda_S^2}
    (\bmta_1\times\bmta_2)_z\nonumber\\
  &&\cdot \bigg[  { \bmk_2\cdot\bmsi_2 \over \omega_2^2} i \bmsi_1 +(1\leftrightarrow2)\nonumber\\
  && + { \bmk_1\cdot\bmsi_1 \bmk_2\cdot\bmsi_2 \over \omega_1^2 \omega_2^2} i (\bmk_1-\bmk_2)\bigg]\,,\label{eq:j2v}`
\end{eqnarray}
where as usual $i=x,y,z$ and $\omega_i=\sqrt{m_\pi^2+ k_i^2}$, etc.
Above we have introduced
\begin{eqnarray}
  h_V(q) &=& \frac{3C^V_+G_E^s(q)\delta_{t't} +C^V_- G_E^v(q) (\tau_z)_{t't}}{\Lambda_S^2}\,,\\
  \tilde h_V(q) &=& \frac{3 C^V_+ G_M^s(q)\delta_{t't} +C^V_- G_M^v(q)(\tau_z)_{t't}}{\Lambda_S^2}\,,
\end{eqnarray}
where $G_E^s(q)$, $G_E^v(q)$, $G_M^s(q)$, and $G_M^v(q)$ are the isoscalar and isovector electric
and magnetic form factors of the nucleons. They are normalized so that
\begin{eqnarray}
  G^s_E(0)&=&G^v_E(0)=1\,,\nonumber\\
  G^s_M(0)&=&1+\kappa_p+\kappa_n\,,\  G^v_M(0)=1+\kappa_p-\kappa_n\,,
\end{eqnarray}
where $\kappa_p$ and $\kappa_n$ are the anomalous magnetic moment of the nucleons. In fact, as explained in
Subsect.~\ref{sec:lecs} we can identify $c_7=\kappa_p+\kappa_n$ and $c_6=\kappa_p-\kappa_n$. 
Some comments are in order. 1) The correction given by the form factors is applied only to the one-body current,
the two-body current being already at N2LO. 2) For simplicity we have used the well-know dipole parametrization
of the form factors, namely we have taken
\begin{eqnarray}
  G^s_X(q)&=& G^p_X(q)+ G^n_X(q)\,,\qquad X=E,M\,,\\
  G^v_X(q)&=& G^p_X(q)- G^n_X(q)\,,\qquad X=E,M\,,\\
  G^p_E(q)&=&G_D(q)\,,\quad G^n_E(q)=-\kappa_n {q^2\over 4M^2} {G_D(q)\over 1+{q^2\over M^2}}\,,\\
  G^p_M(q)&=&(1+\kappa_p) G_D(q)\,,\quad G^n_M(q)=\kappa_n G_D(q)\,,\\
  G_D(q) &=& {1\over (1+{q^2\over\Lambda_V^2})^2} \,,
\end{eqnarray}
where $\Lambda_V=0.84$ GeV has been extracted from fits of  elastic electron scattering data off the proton and
deuteron~\cite{Rocco98,deJager2004}. 3) Usually the form factors are expressed in terms of the
quantity $Q^2=q^2-w^2$, where $w$ is the energy transfer. However, in the present case, $w=(k'^2-k^2)/2M_\chi\ll q$,
so we have assumed $Q\approx q$.  4) As discussed before, we need the form factors at values of $q$ rather small,
where the dipole parametrization is a sufficiently good approximation. 

As usual, we report below the decomposition for the isoscalar operators
\begin{eqnarray}
  (J_{\alpha,\alpha',is}^{(1)})^{\mu}&=& {C^V_+\over \Lambda_S^2} \sum_{\nu=0,2} (J_{\alpha,\alpha',is}^{(1),\nu})^{\mu}\,,\\
  \rho_{\alpha,\alpha',is}^{(1),0} &=&-3G_E^s(q) \delta_{s,s'}\delta_{t,t'}\,,  \label{eq:rho1v1}\\
  \rho_{\alpha,\alpha',is}^{(1),2} &=&-3\big(2G_M^s(q)-G_E^s(q)\big)\delta_{t,t'}\nonumber\\
   &\times&  \bigg[-\frac{q^2}{8M^2}+\frac{i(\bmp'\times\bmp)\cdot\bmsi}{4M^2}\bigg]_{s's}\,, \label{eq:rho1v3}\\
  \bmJ_{\alpha,\alpha',is}^{(1),1} &=&-3 G_E^s(q)  \frac{\bmp+\bmp'}{2M} \delta_{s,s'}\delta_{t,t'}\nonumber\\
  && +3G_M^s(q) \frac{i(\bmq\times\bmsi_{s's})}{2M} \delta_{t,t'}\,,\label{eq:j1v1}\\  
  (J_{\alpha_1,\alpha_1',\alpha_2,\alpha_2',is}^{(2)})^{\mu}&=& 0\,,
\end{eqnarray}
while $\rho_{\alpha,\alpha',is}^{(1),1}=\bmJ_{\alpha,\alpha',is}^{(1),0}=\bmJ_{\alpha,\alpha',is}^{(1),2}=0$. 
Here, we have adopted the notation $J^\mu=(\rho,\bmJ)$. 

\subsection{Axial Interaction}\label{axinteract}
In order to describe the axial interaction we consider the chiral Lagrangian in SU(3) space, which reads~\cite{su3},
\begin{alignat}{2}\label{su3lag}
  \mathcal{L}_{int}&=Tr\big[ \bar{B}\left( i\gamma_\mu D_\mu - M_0\right)B-
    \frac{F}{2}\bar{B}\gamma^\mu \gamma_5\left[u_\mu,B\right]\nonumber\\
  &+\frac{D}{2}\bar{B}\gamma^\mu\gamma^5\left\lbrace u_\mu,B \right\rbrace \big]
   +\frac{f_{M}^{2}}{4}
   \left\langle\nabla_\mu U \nabla^\mu U^\dagger \right\rangle \dots \,.
\end{alignat}
where $D$, $F$, and $f_{M}$ are LECs and the dots stands for other contributions not relevant for this study.
The quantities $B$, $U$, and $u_\mu$ are now $3\times3$ matrices of the various baryon and meson fields
\begin{equation}
B=\left(\begin{array}{ccc} 
         {\Sigma^0\over\sqrt{2}}+{\Lambda\over\sqrt{6}} & \Sigma^+ & p \\
         \Sigma^- & -{\Sigma^0\over\sqrt{2}}+{\Lambda\over\sqrt{6}} & n \\
         \Xi^- & \Xi^0 & -2 {\Lambda\over\sqrt{6}} \\
         \end{array}\right)\,,
\end{equation}
and $U=e^{i\Phi/f_M}$ where
\begin{equation}
\Phi=\left(\begin{array}{ccc} 
         {\pi^0\over\sqrt{2}}+{\eta^8\over\sqrt{6}} & \pi^+ & K^+ \\
         \pi^- & -{\pi^0\over\sqrt{2}}+{\eta^8\over\sqrt{6}} & K^0 \\
         K^- & \overline K^0 & -2 {\eta^8\over\sqrt{6}} \\
         \end{array}\right)\,.
\end{equation}
Moreover, $u=\sqrt{U}$ as usual, and
\begin{equation}
  u_\mu = i\big[ u^\dag \partial_\mu u - u \partial_\mu u^\dag -i u^\dag a_\mu u - i u a_\mu u^\dag\big]\,,
\end{equation}
ad $a_\mu$ is the SU(3) axial current given in Eq.~(\ref{eq:axialsu3}).

We can again expand the Lagrangian in the mesons field $\Phi$. Since we expect that the coupling constants
$C_{A\pm}$ to be small, we will neglect the terms of the expansion that are quadratic in $a_\mu$. Developing the traces
(and retaining only the terms involving nucleons and pions) we get,
\begin{eqnarray}
  \mathcal{L}^A_{int}&=&
  (D+F)\bar{N}\gamma^\mu\gamma^5\tau_zN a_\mu^{(3)}+
  (3F-D) \bar{N}\gamma^\mu\gamma^5N {a_\mu^{(8)}\over\sqrt{3}} \nonumber\\
   &-&2f_\pi \partial^\mu \pi_z   a_\mu^{(3)}
   +{1\over f_\pi} \bar{N} \gamma^\mu (\vec\tau\times\vec\pi)_z N  a_\mu^{(3)}+\cdots
   \,.\label{eq:axlag2}
\end{eqnarray}
In these terms we have identified $f_M=f_\pi$. 
The Hamiltonian density can be obtained using the Legendre transformation, but a particular attention has to be paid to the last term.
The interaction term appearing finally in the Hamiltonian after the transformation reads
\begin{equation}
  \mathcal{H}^A_{int}=\cdots-{1\over 2 f_\pi}\bar{N} \gamma^0 (\vec\tau\times\vec\pi)_z N  a_0^{(3)}
  -{1\over f_\pi}\bar{N} \gamma^i (\vec\tau\times\vec\pi)_z N  a_i^{(3)}+\cdots\,.  \label{eq:axlag2}
\end{equation}

Now $J^{(1)}$, $J^{(2)}$, and $L$ are again four vectors, the chiral order of their
``time'' and ``space'' parts being different. Explicitly, the WIMP current in this case is given by
\begin{equation}
  L_{\bmk' r',\bmk r}^\mu = \bar u^{\chi}_{\bmk' r'}\gamma^\mu\gamma^5 u^{\chi}_{\bmk r}\,,\label{eq:LA}
\end{equation}
which expanded up to order $Q^2$ is given by
\begin{eqnarray}
  (L_{\bmk' r',\bmk r})^{\mu=0}&=&\left(\frac{(\bmk'+\bmk)\cdot\bmsi}{2M_\chi}\right)_{r'r}\,,\\
  (L_{\bmk' r',\bmk r})^{\mu=i}&=&\bigg[\sigma_i-\frac{(\bmk'+\bmk)^2\sigma_i}{8M^2_\chi}\nonumber\\
  &&+\frac{1}{4M^2_\chi}\Bigl(k'_i(\bmsi\cdot\bmk)+k_i(\bmsi\cdot\bmk')\nonumber\\
  &&\qquad -i(\bmk'\times \bmk)_i\Bigr)\bigg]_{r'r}\,.
\end{eqnarray}

\begin{figure}[h!]
  \centering
  \includegraphics[scale=0.5]{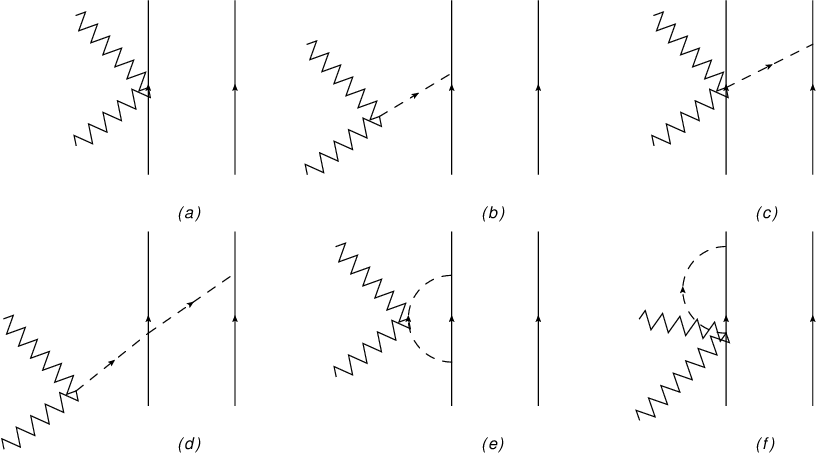}
  \caption{The same as in Fig.~\protect\ref{fig:scalar} but for the axial
  interaction. }
  \label{fig:axial}
\end{figure}

The relevant diagrams for the axial interaction are reported in Fig.~\ref{fig:axial}.
The WIMP-nucleon vertex appearing in the diagram of panel (a)
derives from the interaction terms reported in the first line of Eq.~(\ref{eq:axlag}). 
The minimal order of the space component of $J^{(1)}$
associated to this diagram is $\approx Q^{-3}$, while the time part is $\approx Q^{-2}$. 
Therefore, we include the corrections coming from the expansion of the four-spinors entering the vertex
only in the space component. The contributions of the diagram in panel (b) (the ``pion-pole'' diagram)
behaves analogously. In this case, only isovector transition operators are obtained. 

The diagrams (c) and (d) give the contribution of the one-pion exchanges. The time part of these diagrams is
of order $Q^{-1}$, so we will take into account it. The $WWNN\pi$ vertex derives from the interaction
Hamiltonian given in Eq.~(\ref{eq:axlag2}). On the other hand the spatial parts of the corresponding $J^{(2)}$
are of order $Q^{0}$, so they will be neglected here. Finally, the diagrams in panels (e) and (f)
(and many others) contribute to the axial form factor of the nucleon. These two gives a pure isovector
contribution, so we will take into account them (and many others) by including in the isovector part
of the current the phenomenological axial form factor. The corresponding isoscalar part is not well know, so we will not
include it (in any case it appears at least at the ${\cal O}(Q^0$ order). 
The final expression of the transition densities we will consider is therefore

\begin{eqnarray}
  (J^{(1)}_{\a,\a'})^{\mu=0}&\!\!=\!\!&\!\! - h_A(q) \left(\frac{\bmK\cdot\bmsi}{M}\right)_{s's}\nonumber\\
  &&\!\! -G_A(q) {C^A_-\over \Lambda_S^2} {(\tau_z)_{t't}\over \omega_q^2} \left(\frac{\bmK\cdot\bmsi}{M}\right)_{s's}
  \,,\label{eq:axial0}\\
  (J^{(1)}_{\a,\a'})^{\mu=i}&\!\!=\!\!& \!\! -h_A(q) 
  \bigg[\sigma_i-\frac{K^2\sigma_i}{2M^2}+\frac{1}{4M^2}\Big(p'_i(\bmsi\cdot\bmp)\nonumber\\
  && \qquad + p_i (\bmsi\cdot\bmp')- i(\bmp'\times\bmp)_i\Big)\bigg]_{s's}\nonumber\\
  &&+ G_A(q) {C^A_-\over \Lambda_S^2} {(\tau_z)_{t't}\over \omega_q^2} q^i \bigg[ (\bmq\cdot\bmsi)_{s's} \nonumber\\
    && +{1\over 4M^2}\big(2 \bmK\cdot\bmq \bmK\cdot\bmsi -2 K^2 \bmq\cdot\bmsi\nonumber\\
    && \qquad -{1\over2} q^2 \bmq\cdot\bmsi\big)\bigg]_{s's}\,,\label{eq:axial1}\\
    (J^{(2)}_{\a_2,\a_2'\a_1,\a_1'})^{\mu=0}&=& {g_A^2\over 2 f_\pi^2} {C^A_-\over \Lambda_S^2}
    { (\vec\tau_1\times\vec\tau_2)_z \over \omega_2^2} i \bmk_2\cdot\bmsi_2\,,    \label{eq:axial2}\\
    (J^{(2)}_{\a_2,\a_2'\a_1,\a_1'})^{\mu=i}&=&0\,.\label{eq:axial3}
 \end{eqnarray}
where above $\bmK=(\bmp+\bmp')/2$ and
\begin{eqnarray}
  h_A(q)&=& \frac{(3F-D) C^A_+\delta_{t't}+ G_A(q) C^A_-(\tau_z)_{t't}}{\Lambda_S^2}\,,\\
  G_A(q) &=& {g_A\over (1+ {q^2\over\Lambda_A^2})^2}\,.\label{eq:axialff}
\end{eqnarray}
Note that $D+F\equiv g_A$. Here we assume $\Lambda_A=1$ GeV, as determined from  an analysis of pion electroproduction and neutrino scattering data~\cite{Bodek08,Megias20}
(again, we can safely assume that $Q\approx q$).  Uncertainties in the value of $\Lambda_A$ does not significantly impact
predictions for the WIMP cross-section as $q\ll \Lambda_A$. In the pion-pole contribution, 
usually it should appear the pseudoscalar form factor $G_{PS}(q)$. From our chiral analysis, we obtain $G_{PS}(q)=G_A(q)/(m_\pi^2+q^2)$ (the pole contribution), well verified by the experimental data~\cite{Kammel18}.

As usual, we report below the decomposition of the isoscalar operators
\begin{eqnarray}
  (J_{\alpha,\alpha',is}^{(1)})^{\mu}&=& {C^A_+\over \Lambda_S^2} \sum_{\nu=0,2} (J_{\alpha,\alpha',is}^{(1),\nu})^{\mu}\,,\\
  \rho_{\alpha,\alpha',is}^{(1),1}  &=& -(3F-D) \left(\frac{\bmK\cdot\bmsi}{M}\right)_{s's} \delta_{t't} \,, \label{eq:rho1a1}\\
  \bmJ_{\alpha,\alpha',is}^{(1),0}  &=& -(3F-D) \bmsi_{s's} \delta_{t't}\,, \label{eq:j1a0}\\
  \bmJ_{\alpha,\alpha',is}^{(1),2}  &=& -(3F-D) 
  \bigg[-\frac{K^2\bmsi}{2M^2}+{1\over 4M^2}\Big(\bmp'(\bmsi\cdot\bmp)\nonumber\\
  &&+ \bmp (\bmsi\cdot\bmp')- i(\bmp'\times\bmp)\Big)\bigg]_{s's}\delta_{t't}\,,\label{eq:j1a2}\\
  (J_{\alpha_1,\alpha_1',\alpha_2,\alpha_2',is}^{(2)})^{\mu}&=& 0\,,
\end{eqnarray}
while $\rho_{\alpha,\alpha',is}^{(1),0}=\rho_{\alpha,\alpha',is}^{(1),2}=\bmJ_{\alpha,\alpha',is}^{(1),1}=0$.
Again $J^\mu=(\rho,\bmJ)$.

\subsection{Tensor interaction}\label{tensinteract}

In the case of tensor interaction it is necessary to start from the construction of the nucleon Lagrangian where the
current $t^{\mu \nu}$ appears (to be noticed that such terms are thought to be rather suppressed~\cite{Bishara17}). The hadronic Lagrangian for a tensor current has been constructed in Refs.~\cite{Cata07,Magid23,Liang24}, however, here we will briefly recall the steps. We have previously seen that the quark Lagrangian with external tensor current reads
\begin{equation}
  \mathcal{L}^{tens}_{q}=\bar{q}\sigma_{\mu\nu}t^{\mu\nu}q\,\,.
\end{equation}
Since $t^{\mu\nu}=t^{\mu\nu}\,^{\dagger}$, this term can be rewritten as
\begin{equation}
  \mathcal{L}^{tens}_{q}=\bar{q}_R\sigma_{\mu\nu}t^{\mu\nu}q_L+\bar{q}_L\sigma_{\mu\nu}t^{\mu\nu\dagger}q_R.
\end{equation}
Assuming that $t^{\mu\nu}$ transforms under chiral transformations as
\begin{eqnarray}
  t^{\mu\nu}\rightarrow R t^{\mu\nu}L^\dagger\,,\\
  t^{\mu\nu\dagger}\rightarrow L t^{\mu\nu\dagger}R^\dagger\,,
\end{eqnarray}
where $L$ ($R$) represents a local rotation in the isospin space of the left (right) components,
and remembering that under these transformations the nucleon field $N$ (a doublet in isospin space)
and the pionic unitary matrix $U=e^{i\vec\pi(x)\cdot\tau/f_\pi}$ transform as 
\begin{eqnarray}
  N&\rightarrow &hN\,,\\
  U&\rightarrow &RuL^\dagger\,,\\
  u&\rightarrow &Ruh^\dagger=huL^\dagger\,,
\end{eqnarray}
where $u=\sqrt{U}$ and $h$ is a $SU(2)$ matrix depending in a complicate way on $L$, $R$, and $\vec\pi(x)$,
it can be seen that Lagrangian terms invariant under chiral transformations to the lowest order are
$\bar{N}\sigma_{\mu\nu}T^{\mu\nu}_\pm N$, where
\begin{equation}
  T^{\mu\nu}_\pm =  u t^{\mu\nu\dagger}u\pm u^\dagger t^{\mu\nu}u^\dagger\,.
\end{equation}
  In fact, it is easy to prove that,
\begin{equation}
  T^{\mu\nu}_\pm \rightarrow h T^{\mu\nu}_\pm h^\dagger\label{Operf}.
\end{equation}
Considering that among the two operators~(\ref{Operf}) only $T^{\mu\nu}_+$ is invariant under parity,
charge and hermitian conjugation, we obtain that the lowest order Lagrangian of the nucleons will be
\begin{equation}
  \mathcal{L}^T_{int}= \tilde{c}_1 \bar{N}\sigma_{\mu\nu}\langle T^{\mu\nu}_+
  \rangle N+\tilde{c}_2 \bar{N}\sigma_{\mu\nu}\hat{T}^{\mu\nu}_+ N \,,
\end{equation}
where $\tilde{c}_1$ and $\tilde{c}_2$ are new LECs.
Higher order terms can be constructed combining $T^{\mu\nu}_\pm$ with $u_\mu$, ect.
Here for simplicity we consider only the lowest order term given above.

Expanding $T^{\mu\nu}_+$ in power of the pion fields and considering only the lowest order terms,
the nucleon-WIMP interaction Lagrangian becomes
\begin{equation}
  \mathcal{L}^{T}_{int}=\bar{N}\sigma_{\mu\nu}\frac{1}{\Lambda_S^2}
  \left(4\tilde{c}_1C^T_++2\tilde{c}_2C^T_-\tau_z\right) N \bar{\chi}\sigma^{\mu\nu}\chi+\ldots\,.\label{eq:telag2}
\end{equation}
We remember that in this case the nucleonic and WIMP currents are four-tensors, 
and clearly only the off-diagonal elements are different from zero. 
The WIMP tensor in this case is given by
\begin{equation}
  L_{\bmk' r',\bmk r}^{\mu\nu} = \bar u^{\chi}_{\bmk' r'}\sigma^{\mu\nu} u^{\chi}_{\bmk r}\,,\label{eq:LT}
\end{equation}
which expanded up to order $Q^2$ reads
\begin{eqnarray}
  (L_{\bmk' r',\bmk r})^{0i}&=&\bigg( \frac{iq_i}{2M_\chi}-\frac{(\bmQ\times\bmsi)_i}{2M_\chi}\bigg)_{r'r}\,,\nonumber\\
  (L_{\bmk' r',\bmk r})^{ij}&=&\epsilon_{ij\ell}\bigg(\sigma_\ell-\sigma_\ell\frac{q^2}{8M_\chi^2}+
  q_\ell\frac{(\bmq\cdot\bmsi)}{8M_\chi^2}\nonumber\\
  && -Q_\ell\frac{(\bmQ\cdot\bmsi)}{2M_\chi^2}-i\frac{(\bmq\times\bmQ)_\ell}{4M_\chi^2}\bigg)_{r'r}\,,
\end{eqnarray}
where $\bmQ=(\bmk+\bmk')/2$. 

\begin{figure}[h!]
  \centering
  \includegraphics[scale=0.5]{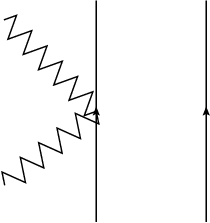}
  \caption{The same as in Fig.~\protect\ref{fig:scalar} but for the tensor
  interaction. }
  \label{fig:tensor}
\end{figure}

The only diagram we consider here for the tensor interaction is that reported in Fig.~\ref{fig:tensor}.
The WIMP-nucleon vertex appearing in the diagram of panel (a)
derives from the interaction terms reported in Eq.~(\ref{eq:telag}).
The minimal order of the time-space component of $J^{(1)}$
associated to this diagram is $\approx Q^{-2}$, while the space-space part is $\approx Q^{-3}$. Since $L^0$
is of order $Q$, the product $J_0^{(1)}L_0$ is nominally of order $Q^{-1}$. 
Therefore, we include the corrections coming from the expansion of the four-spinors entering the vertex
only in the space-space component.

The final expressions for the antisymmetric single nucleon operator $(J^{(1)}_{\a',\a})^{\mu\nu}$ can be written in terms of two current vectors as
\begin{equation}
    (J^{(1)}_{\a,\a'})^{ij}= \epsilon_{ijl} (J^{(A)}_{\a,\a'})^{l}\,, \quad
  (J^{(1)}_{\a,\a'})^{0i}= (J^{(B)}_{\a,\a'})^{i}\,,
\end{equation}
where
\begin{eqnarray}
  \bmJ^{(A)}_{\a,\a'}&=& \bigg(4\tilde{c}_1\frac{C^T_+}{\Lambda_S^2}\delta_{t't}
  +2\tilde{c}_2\frac{C^T_-}{\Lambda_S^2}(\tau_z)_{t't}\bigg)\nonumber\\
  &&\times\bigg(\bmsi-\bmsi\frac{q^2}{8M^2}
  +\bmq \frac{(\bmsi\cdot\bmq)}{8M^2}\nonumber\\
  &&\quad -\bmK\frac{(\bmsi\cdot\bmK)}{2M^2}+\frac{i(\bmq\times\bmK)}{4M^2}\bigg)_{s's}\,,\label{eq:j1teij}\\
  \bmJ^{(B)}_{\a,\a'}&=&\left(4\tilde{c}_1\frac{C^T_+}{\Lambda_S^2}\delta_{t't}
  +2\tilde{c}_2\frac{C^T_-}{\Lambda_S^2}(\tau_z)_{t't}\right)\nonumber\\
  &&\times\bigg(-\frac{i\bmq}{2M}-\frac{\bmK\times\bmsi)}{M}\bigg)_{s's}\,.\label{eq:j1te0i}
\end{eqnarray}

As usual, we report below the decomposition of the isoscalar operators
\begin{eqnarray}
  \bmJ^{(A)}_{\a,\a',is}&=& \left(\frac{C^T_+}{\Lambda_S^2}\right)
   \sum_{\nu=0,2} \bmJ_{\alpha,\alpha',is}^{(A),\nu}\,,\\
  \bmJ_{\alpha,\alpha',is}^{(A),0} &=&4\tilde{c}_1 (\bmsi)_{s's}\delta_{t't}\,,\label{eq:jta1}\\
  \bmJ_{\alpha,\alpha',is}^{(A),2} &=&4\tilde{c}_1 \bigg(-\bmsi\frac{q^2}{8M^2}
    +\bmq \frac{(\bmsi\cdot\bmq)}{8M^2}\nonumber\\
  && -\bmK\frac{(\bmsi\cdot\bmK)}{2M^2}+\frac{i(\bmq\times\bmK)}{4M^2}\bigg)_{s's}\delta_{t't}\,,\label{eq:jta2}\\
      \bmJ^{(B)}_{\a,\a',is}&=&\left(\frac{C^T_+}{\Lambda_S^2}\right)
  \sum_{\nu=0,2}\bmJ_{\alpha,\alpha',is}^{(B),\nu}\,,\\
  \bmJ_{\alpha,\alpha',is}^{(B),1}&=&
  4\tilde{c}_1 \bigg(-\frac{i\bmq}{2M}-\frac{\bmK\times\bmsi}{M}\bigg)_{s's}\delta_{t't}\,,\label{eq:jtb1}
\end{eqnarray}
while $\bmJ_{\alpha,\alpha',is}^{(A),1}=\bmJ_{\alpha,\alpha',is}^{(B),0}=\bmJ_{\alpha,\alpha',is}^{(B),2}=0$.

\subsection{Values of the LECs}\label{sec:lecs}
We report the values of the LECs entering our calculation in Table~\ref{tab:lecs}.
The value of $B_c$ is related to the quark condensate in vacuum. We have already discussed that
this parameter is also related to the pion mass. In fact, expanding the chiral Lagrangian
in terms of the pion field, and looking to the terms proportional
to the pion field square, we can identify $m_\pi^2=2 m_q B_c$, where $m_q$ is the average
between the mass of $u$ and $d$ quarks. Adopting the value $m_q=3.45$ MeV~\cite{param},
we can estimate $B_c\approx 2.78$ GeV. Note that the relation between $m_\pi^2$ and $B_c$
will have higher order corrections coming from $\mathcal{L}_{\pi}^{(4)}$, etc. Other
estimates of $B_c$ come from the Gell-Mann-Oakes-Rennes relation between the mass of
pseudoscalar mesons~\cite{gellmann}, or directly from Lattice calculations. The more
precise estimate obtained for $B_c$ using the latter method is  
$B_c=2.40\pm 0.03$ GeV~\cite{aoki} (see also Ref.~\cite{wang} for a more recent estimate).
In our work we have adopted the value of Ref.~\cite{aoki}, representing an average of the
results of different lattice calculations.

The scalar current has been written in terms of $\sigma_{\pi N}$ constant. Here we will assume $\sigma_\pi=-59.1\pm 3.5$ MeV, as extracted from a Roy-Steiner analysis of pion-nucleon scattering~\cite{Hofe15} (in the following, we do not take into account of the small associated error). 

The LECs $d_{18}$ and $d_{19}$ entering the pseudoscalar Lagrangian are not well known. Since the pseudoscalar
interaction will produce a very small reaction rate, we will take $d_{18}+2d_{19}=1$ GeV$^{-2}$,
namely a sort ``natural'' value.

For the vector current, the LECs $c_6$ and $c_7$ are simply related to the anomalous magnetic moment of the nucleons. In fact,
assuming that the vector external current
is given by the electromagnetic field, then the one-body nuclear current would be given by
\begin{alignat}{2}
 (J^{(1)}_{\alpha,\alpha'})^{\mu=i}_{\rm EM}&= {(\bmp+\bmp')_i\over 2M}\frac{(1+\tau_z)}{2}-
  i {(\bmq\times\bmsi)_i\over 2M}\nonumber\\&\cdot\left( {1+\tau_z\over 2} +
  {c_7+c_6\tau_z\over 2}\right)\ ,
\end{alignat}
from which we can identify~\cite{Rocco98}
\begin{equation}
 c_6=\kappa_p-\kappa_n\ , \qquad
 c_7=\kappa_p+\kappa_n\ ,
\end{equation}
where $\kappa_p=1.793$ ($\kappa_n=-1.913$) are the anomalous magnetic moments of the proton (neutron)
in unit of the nuclear magneton. The values reported in Table~\ref{tab:lecs} are obtained from these relations.

The values for the LECs $D$ and $F$, which enter the axial current in the SU(3) formalism,  are taken from
Ref.~\cite{dfsu3}. Note that $F+D\approx 1.26\approx g_A$~\cite{dfsu3}.

Finally, the values of the LECs $\tilde c_1$ and $\tilde c_2$ entering the tensor case can be obtained from the results of a recent lattice calculation on the tensor charges of the nucleons~\cite{alexandrou}. 
We have found
\begin{equation}
  4\tilde c_1 \equiv g_T^{u+d} = 0.582\pm 0.016\ ,\quad
  2\tilde c_2 \equiv g_T^{u-d} = 1.004\pm 0.021\ ,
\end{equation}
where the quantities $g_T^{u\pm d}$ were calculated in Ref.~\cite{alexandrou}.

As already stated, the values of the LECs used in this work are summarized in table~\ref{tab:lecs}. Note that here we have not tried to quantify the propagation of the error with which these LECs are known to the DM rates. This task will be demanded to a successive work. 

\begin{table}[htbp]
\centering
\begin{tabular}{l|c}
\hline
LEC & Value\\
\hline
$g_A$ & $\phantom{-}1.27\phantom{\,\mathrm{MeV}^{-1}}$\\
$f_\pi$ & $\phantom{-}92.4\,\mathrm{MeV}\phantom{^{-1}}$\\
$B_c$&$\phantom{-}2.40\,\mathrm{GeV}\phantom{^{-1}}$\\
$c_6$&$\phantom{-}3.71\,\phantom{\mathrm{GeV}^{-2}}$\\
$c_7$&$-0.12\,\phantom{\mathrm{GeV}^{-2}}$\\
$d_{18}+2d_{19}$&$\phantom{-}1.00\,\mathrm{GeV}^{-2}$\\ 
$D$&$\phantom{-}0.86\phantom{\,\mathrm{GeV}^{-2}}$\\
$F$&$\phantom{-}0.39\phantom{\,\mathrm{GeV}^{-2}}$\\
$4\tilde{c_1}$&$\phantom{-}0.58\,\phantom{\mathrm{GeV}^{-2}}$\\
\hline
\end{tabular}
\caption{LECs values used in this work.}
\label{tab:lecs}
\end{table}

\section{The quantities $I(a,\bm{b},c,\bm{d})$}\label{app:b}

In this Appendix we list the quantities $I(a,\bm{b},c,\bm{d})$ entering the expression of the rate~(\ref{eq:d3rate})
for the various cases.

\begin{itemize}
\item Scalar interaction
\begin{equation}
  I^S_1 = I_1\big(1-V^2-{\bmV\cdot\bmq\over M_\chi}-{q^2\over 4M_\chi^2}, 2\bmV+{\bmq\over M_\chi}, -1 , \bm{0}\big) \,.
\end{equation}

\item Pseudoscalar interaction
\begin{equation}
  I^P_1 = I\big({q^2\over 4M_\chi^2}, \bm{0} , 0 , \bm{0}\big) \,.
\end{equation}
\newpage
\item Vector interaction

  \begin{eqnarray}
  I^V_1 &=& I\big(1-{q^2\over 4M_\chi^2}, \bm{0} , 0 , \bm{0}\big)\,,  \nonumber \\
  I^V_2 &=& I\big(-\bmV\cdot\hat q-{q\over 2 M_\chi}, \hat q , 0 , \bm{0}\big)\,,   \\
  I^V_3 &=& I\big((\bmV\cdot\hat q)^2+{\bmV\cdot\bmq\over M_\chi} + {q^2\over 4 M_\chi^2},
                           -2\hat q (\bmV\cdot\hat q+{q\over 2 M_\chi}), 0 , \hat q \big)\,,  \nonumber \\
  I^V_4 &=& I\big( {1\over2}(V^2- (\bmV\cdot\hat q)^2)+{q^2\over 4M_\chi^2}, -\bmV+\hat q \bmV\cdot\hat q, \nonumber\\
        &&\qquad                   {1\over 2}, i{\hat q\over\sqrt{2}}\big)\,.\nonumber
  \end{eqnarray}
  
\item Axial interaction
  \begin{eqnarray}  
  I^A_1 &=& I\big(V^2+{\bmV\cdot\bmq\over M_\chi}+{q^2\over 4M_\chi^2}, -2\bmV-{\bmq\over M_\chi} , 1 , \bm{0}\bigg)\,, \nonumber \\
  I^A_2 &=& I\big(-\bmV\cdot\hat q-{q\over 2 M_\chi}, \hat q, 0, \bm{0} \big)\,, \\
  I^A_3 &=& I\big( 1+(\bmV\cdot\hat q)^2 -V^2- {q^2\over 4M_\chi^2}, 2 \bmV -2 \hat q (\bmV\cdot\hat q),
                           -1, \hat q \big)\,,  \nonumber \\
  I^A_4 &=& I\big( 1-{1\over2}V^2- {1\over2}(\bmV\cdot\hat q)^2-{\bmV\cdot\bmq \over M_\chi}-{q^2\over 4M_\chi^2},\nonumber\\
        &&\qquad          \bmV+\hat q \bmV\cdot\hat q+{\bmq\over M_\chi},
                           -{1\over 2}, i{\hat q\over\sqrt{2}}\big)\,.\nonumber
\end{eqnarray}

\item Tensor interaction
\begin{eqnarray}
  I^{T,A}_3 &=&  I\big(1-(\bmV\cdot\hat q)^2-{\bmV\cdot\bmq\over M_\chi} -{q^2\over 4 M_\chi^2},\nonumber\\
  &&\qquad 2 \hat q (\bmV\cdot\hat q +{q\over 2M_\chi}), 0, i\hat q\big)\,, \\
   I^{T,A}_4 &=&  I\big(1-{1\over2}V^2+{1\over2}(\bmV\cdot\hat q)^2-{q^2\over 4 M_\chi^2},\nonumber\\
   && \qquad \bmV-\hat q (\bmV\cdot\hat q), -{1\over2},  {\hat q\over\sqrt{2}} \big)\,,\\ 
   I^{T,B}_3 &=&  I\big(V^2-(\bmV\cdot\hat q)^2+{q^2\over 4 M_\chi^2},\nonumber\\
   &&\qquad -2\bmV+2\hat q (\bmV\cdot\hat q), 1 , i\hat q \big)\,,\\ 
   I^{T,B}_4 &=&  I\big({V^2\over 2} +{(\bmV\cdot\hat q)^2\over2}+{\bmV\cdot\bmq\over M_\chi}+{q^2\over 4 M_\chi^2},\nonumber\\
   && \qquad -\bmV-\hat q (\bmV\cdot\hat q)-{\bmq\over M_\chi},{1\over2}, {\hat q\over\sqrt{2}} \big)\,,\\
   I^{T,AB}_4 &=& I\big(-\bmV\cdot\hat\bmq -{q\over 2M_\chi}, \hat\bmq,0,0\big)\,.\nonumber
\end{eqnarray}

\end{itemize}



\begin{thebibliography}{0}%
\makeatletter
\providecommand \@ifxundefined [1]{%
 \@ifx{#1\undefined}
}%
\providecommand \@ifnum [1]{%
 \ifnum #1\expandafter \@firstoftwo
 \else \expandafter \@secondoftwo
 \fi
}%
\providecommand \@ifx [1]{%
 \ifx #1\expandafter \@firstoftwo
 \else \expandafter \@secondoftwo
 \fi
}%
\providecommand \natexlab [1]{#1}%
\providecommand \enquote  [1]{``#1''}%
\providecommand \bibnamefont  [1]{#1}%
\providecommand \bibfnamefont [1]{#1}%
\providecommand \citenamefont [1]{#1}%
\providecommand \href@noop [0]{\@secondoftwo}%
\providecommand \href [0]{\begingroup \@sanitize@url \@href}%
\providecommand \@href[1]{\@@startlink{#1}\@@href}%
\providecommand \@@href[1]{\endgroup#1\@@endlink}%
\providecommand \@sanitize@url [0]{\catcode `\\12\catcode `\$12\catcode `\&12\catcode `\#12\catcode `\^12\catcode `\_12\catcode `\%12\relax}%
\providecommand \@@startlink[1]{}%
\providecommand \@@endlink[0]{}%
\providecommand \url  [0]{\begingroup\@sanitize@url \@url }%
\providecommand \@url [1]{\endgroup\@href {#1}{\urlprefix }}%
\providecommand \urlprefix  [0]{URL }%
\providecommand \Eprint [0]{\href }%
\providecommand \doibase [0]{https://doi.org/}%
\providecommand \selectlanguage [0]{\@gobble}%
\providecommand \bibinfo  [0]{\@secondoftwo}%
\providecommand \bibfield  [0]{\@secondoftwo}%
\providecommand \translation [1]{[#1]}%
\providecommand \BibitemOpen [0]{}%
\providecommand \bibitemStop [0]{}%
\providecommand \bibitemNoStop [0]{.\EOS\space}%
\providecommand \EOS [0]{\spacefactor3000\relax}%
\providecommand \BibitemShut  [1]{\csname bibitem#1\endcsname}%
\let\auto@bib@innerbib\@empty
\end{thebibliography}%


\begin{thebibliography}{100}
%
\bibitem{bertone} See for example, G. Bertone {\it et al.},"A New Era in the Quest for Dark Matter", JCAP {\bf 1803}, 026 (2018),{\tt arXiv:1810:01668}, and ref. therein.
%
\bibitem{Feng23} J. L. Feng, "The WIMP Paradigm: Theme and Variations", SciPost Phys. Lect. Notes 71 (2023) {\tt https://doi.org/10.21468/SciPostPhysLectNotes.71}
%
\bibitem{Bottaro22} S. Bottaro {\it et al.}, "Closing the window on WIMP Dark Matter", Eur. Phys.J. C {\bf 82}, 1 (2022); "The last Complex WIMPs standing", {\tt arXiv:2205:04486}
%
\bibitem{FIP23} C. Antel {\it et al.}, "Feebly Interacting Particles: FIPs 2022 workshop report", {\tt arXiv:2305.01715}
%
\bibitem{Baudis} L. Baudis, "Dark matter detection", J. Phys. G {\bf 43}, 044001 (2016).
%
\bibitem{Goodman_2010} J. Goodman, M. Ibe, A. Rajaraman, W. Shepherd, T. M. P. Tait, and H. B.  Yu,  ''Constraints on dark matter from colliders'', Phys.  Rev.  D {\bf 82}, 116010 (2010)
%
\bibitem{Goodman_2011} J. Goodman, M. Ibe, A. Rajaraman, W. Shepherd, T. M. P. Tait, "Constraints on Light Majorana dark Matter from Colliders", Phys.Lett.B {\bf 695}, 185-188 (2011) 
%
\bibitem{Cooley} J. Cooley, "Dark Matter direct detection of classical WIMPs ", SciPost Phys. Lect. Notes 55 (2022) {\tt https://scipost.org/SciPostPhysLectNotes.55}
%
\bibitem{relicdensity} G.Bertone {\it et al.}, "Particle Dark Matter: Evidence, Candidates and Constraints", Phys.Rept, {\bf 405},279-390 (2005), {\tt[hep-ph/0404175]}.
%
\bibitem{Weinberg68} S. Weinberg,  ''Nonlinear realizations of chiral symmetry'', Phys. Rev. {\bf 166}, 1568, (1968)
%
\bibitem{GL84} J. Gasser and H. Leutwyler,  ''Chiral Perturbation Theory to One Loop'', Annals of Phys. {\bf 158}, 142 (1984)
%
\bibitem{Scherer86} S. Scherer, "Introduction to chiral perturbation theory", in {\it Advances in Nuclear Physics, Vol. 27}, edited by J. W. Negele and E. W. Vogt (Kluwer Academic/Plenum Publishers , New York, 1986).
%
\bibitem{Weinberg90} S.\ Weinberg,
"Nuclear forces from chiral Lagrangians", Phys.\ Lett.\  B {\bf 251}, 288 (1990);
"Effective chiral Lagrangians for nucleon-pion interactions and nuclear forces ", Nucl.\ Phys.\  B {\bf 363}, 3 (1991);
"Three-body interactions among nucleons and pions", Phys.\ Lett.\  B {\bf 295}, 114 (1992)
%
\bibitem{Bernard95} V. Bernard, N. Kaiser, and U.-G. Meissner, ''Chiral dynamics in nucleons and nuclei'', Int. J. Mod. Phys. E {\bf 4}, 193, (1995)
%
%

%
%
%
%
\bibitem{Klos13} P. Klos, J. Menéndez, D. Gazit, and A. Schwenk, 
"Large-scale nuclear structure calculations for spin-dependent WIMP scattering with chiral effective field theory currents", Phys.  Rev.  D {\bf 88}, 083516 (2013)
%
\bibitem{Bishara17} F. Bishara, J. Brod, B. Grinstein, J. Zupan, "Chiral Effective Theory of Dark Matter Direct Detection", JCAP {\bf 02}, 009 (2017) 
%
\bibitem{Gazda17} D. Gazda, R. Catena, and C. Forss\'en, "Ab initio nuclear response functions for dark matter searches", Phys. Rev. D {\bf 95}, 103011 (2017)
%
\bibitem{Korber17} C. K\"orber, A. Nogga, and J. de Vries, "First-principle calculations of dark matter scattering off light nuclei", Phys. Rev. C {\bf 96}, 035805 (2017)
%
\bibitem{Fieguth18} A. Fieguth, M. Hoferichter, P. Klos, J. Men\'endez, A. Schwenk, and C. Weinheimer, "Discriminating WIMP-nucleus response functions in present and future XENON-like direct detection experiments", Phys. Rev. D {\bf 97}, 103532 (2018)
%
\bibitem{Andreoli19} L. Andreoli, V. Cirigliano, S. Gandolfi, and F. Pederiva, "Quantum Monte Carlo calculations of dark matter scattering off light nuclei", Phys. Rev. C {\bf 99}, 025501 (2019)
%
\bibitem{Vries23} J. de Vries,  C. K\"orber, A. Nogga, and S. Shain, " Dark matter scattering off $\heq$ in chiral effective field theory",
{\tt ArXiv:2310.11343} (2023)
%
\bibitem{heliumdm} W. Guo and D. N. McKinsey, "Concept for a dark matter detector using liquid helium-4", Phys. Rev. D {\bf 87}, 115001 (2013).
%
\bibitem{Richardson_2022}T. R. Richardson, X.  Lin, S.  T. Nguyen, ''Large-Nc constraints for elastic dark matter-light nucleus scattering in pionless effective field theory'', Phys. Rev. C {\bf 106}, 044003 (2022)
%
\bibitem{Fitzpatrick13} A. L. Fitzpatrick, W. Haxton, E. Katz,
N. Lubbers, and Y. Xu, “The Effective Field
Theory of Dark Matter Direct Detection,” JCAP {\bf 02}, 004 (2013)
%
\bibitem{Anand14} N. Anand, A. L. Fitzpatrick, and W. C. Haxton, “Weakly interacting massive particle-nucleus elastic scattering response,” Phys. Rev. C {\bf 89}, 065501 (2014)
%
\bibitem{Agnes20}  P. Agnes et al. (DarkSide-50), “Effective field theory interactions for liquid argon target in DarkSide- 50 experiment,” Phys. Rev. D 101, 062002 (2020),
%
\bibitem{Akerib21}  D. S. Akerib et al. (LUX), “Constraints on effective field theory couplings using 311.2 days of LUX data,” Phys. Rev. D 104, 062005 (2021)
%
\bibitem{Jeong22} I.  Jeong, S. Kang, S. Scopel, and G. Tomar, "WimPyDD: An object–oriented Python code for the calculation of WIMP direct detection signals", Comput. Phys. Commun. 276, 108342 (2022) 
%
\bibitem{Heimsoth23} Daniel J. Heimsoth et al., "The uncertainties on the EFT coupling limits for direct dark matter detection experiments stemming from uncertainties of target properties", {\tt arXiv:2305.08991}
%
\bibitem{hoferichter2015ch}  M. Hoferichter, P. Klos, and A. Schwenk, ''Chiral power counting of one- and two-body currents in direct detection of dark matter'',  Phys. Lett. B {\bf 746}, 410 (2015)
%
\bibitem{Aprile23} E. Aprile {\it et al.} (XENON), “Effective Field Theory and
Inelastic Dark Matter Results from XENON1T,” {\tt arXiv:2210.07591}
%
\bibitem{baroni15} A. Baroni, L. Girlanda, S. Pastore, R. Schiavilla, M. Viviani, "Nuclear axial currents in chiral effective field theory",  Phys. Rev. C {\bf{93}}, 015501 (2016); Erratum: Phys. Rev. C {\bf{93}}, 049902 (2016).
%
\bibitem{peskin} See, for example, M. E. Peskin, D. V. Schroeder {\it An Introduction to Quantum Field Theory} ( Westeren Press, 2005).
%
\bibitem{georgi09} H. Georgi, {\it Weak Interactions and Modern Particle Theory}, Dover (2009).
%
\bibitem{gross04} F. Gross, {\it Relativistic Quantum Mechanics and Field Theory}, Wiley - VHC (2004).
%
\bibitem{fettes00} See, for example, N. Fettes, Ulf-G. Meissner, M. Mojzis, S. Steininger, "The Chiral Effective Pion-Nucleon Lagrangian of Order p4",  Ann. Phys. {\bf{283}}, 273 (2000); Erratum-ibid. {\bf 288}, 249 (2001)
%
\bibitem{Rocco98} J. Carlson and R. Schiavilla, "Structure and dynamics of few-nucleon systems",  Rev. Mod. Phys., {\bf 70}, 743, 1998
%
\bibitem{deJager2004} C.E. Hyde-Wright and K. de Jager,"Electromagnetic Form Factors of the Nucleon and Compton Scattering",  Annu. Rev. Nucl. Part. Sci., {\bf 54}, 217 (2004)
%
\bibitem{su3} Jose Antonio Oller, Michela Verbeni, Joaquim Prades, "Meson-Baryon Effective Chiral Lagrangians to $O(q^3)$", JHEP (2006),	{\tt arXiv:hep-ph/0608204}.
%
\bibitem{Bodek08}  A. Bodek, S. Avvakumov, R. Bradford, and H. S. Budd, ''Vector and axial nucleon form factors: A duality constrained parameterization'', Eur. Phys. J. C {\bf 53}, 349–354 (2008)
%
\bibitem{Megias20} G. D. Megias, S. Bolognesi, M. B. Barbaro, and E. Tomasi-Gustafsson, ''New evaluation of the axial nucleon form factor from electron- and neutrino-scattering data and impact on neutrino-nucleus cross-section'', Phys. Rev. C {\bf 101}, 025501 (2020)
%
\bibitem{Kammel18}R. J. Hill, P. Kammel, W. J. Marciano, and A. Sirlin,  ''Nucleon Axial Radius and Muonic Hydrogen - A New Analysis and Review '', Rept.Prog.Phys. {\bf 81}, 096301 (2018)
%
\bibitem{Cata07}  O.  Cata and  V. Mateu, ''Chiral Perturbation Theory with tensor sources'',  	JHEP {\bf 0709}, 078 (2007)
%
\bibitem{Magid23}  A. Glick-Magid, "Non-relativistic nuclear reduction for tensor couplings in dark matter
direct detection and $\mu\rightarrow e$ conversion", {\tt ArXiv:2312.08339} (2023)
%
\bibitem{Liang24} J. H. Liang, Y.  Liao, X. D.  Ma, and H. L.  Wang, ''Comprehensive constraints on fermionic dark matter-quark tensor interactions in direct detection experiments'', {\tt ArXiv:2401.05005} (2024)
%
\bibitem{Hofe15} M. Hoferichter, J. Ruiz de Elvira, B. Kubis, and U.-G. Meißner, "High-Precision Determination of the Pion-Nucleon $\sigma$ Term from Roy-Steiner Equations",  Phys. Rev. Lett. {\bf 115}, 092301 (2015)
%
\bibitem{param} C.Patrignani {\it et al.} (Particle Data Group), {\it Chin.Phys}, {\bf C40}, 100001 (2016).
%
\bibitem{gellmann} M. Gell-Mann, R. J. Oakes and B. Renner, "Behavior of Current Divergences under SU3×SU3",  Phys. Rev. 175, 2195 (1968).
%
\bibitem{aoki} S. Aoki {\it et al.}, "Review of lattice results concerning low-energy particle physics", [FLAG Collaboration],  Eur. Phys. J.
C77, 112 (2017), {\tt arXiv:1607.00299}.
%
\bibitem{wang}C. Wang {\it et al.},"Quark chiral condensate from the overlap quark propagator",   Chin. Phys. C41, 053102 (2017).
%
\bibitem{dfsu3} A. Hosaka, T. Myo, H. Nagahiro, K. Nawa, Hadron and Nuclear Physics 09, Osaka University, Japan, (2009)
%
\bibitem{alexandrou} C. Alexandrou {\it et al.},"Erratum: Nucleon scalar and tensor charges using lattice QCD simulations at the physical value of the pion mass ",  Phys. Rev. D {\bf 96}, 099906 (2017).
%
\bibitem{Potforliq} M.Cadeddu {\it et al.}, "Directional dark matter detection sensitivity of a two-phase liquid argon detector", JCAP 01 (2019) 014, {\tt arXiv:1704.03741 [}.
%
\bibitem{AV18} R.B.\ Wiringa, V.G.J.\ Stoks, and R.\ Schiavilla, "Accurate nucleon-nucleon potential with charge-independence breaking",
	       Phys.\ Rev.\ C {\bf 51}, 38  (1995)
%
\bibitem{MEN17} D.R. Entem, R. Machleidt, and Y. Nosyk, "High-quality two-nucleon potentials up to fifth order of the chiral expansion", 
                Phys. Rev. C {\bf 96}, 024004 (2017)
%
\bibitem{Furn15a}R. J. Furnstahl, D. R. Phillips and S. Wesolowski, ``A recipe for EFT uncertainty quantification in nuclear physics'', J. Phys. G: Nucl. Part. Phys. \textbf{42}, 034028 (2015)
%
\bibitem {Furn15b}R. J. Furnstahl, N. Klco, D. R. Phillips, and S. Wesolowski,  ''Quantifying truncation errors in effective field theory", Phys. Rev. C {\bf 92}, 024005 (2015)
%
\bibitem{Weso16}S. Wesolowski, N. Klco, R. J. Furnstahl, D. R. Phillips, and A. Thapaliya,  ''Bayesian parameter estimation for effective field theories'',
  J. Phys. G: Nucl. Part. Phys. {\bf 43}, 074001 (2016)
%
\bibitem{Weso21} S. Wesolowski {\it et al.}, ``Rigorous constraints on three-nucleon forces in chiral effective field theory from fast and accurate calculations of few-body observables", Phys. Rev. C {\bf 104}, 064001 (2021)
%
\bibitem{Pudliner97} B. S. Pudliner, V.R. Pandharipande, J. Carlson, S.C. Pieper, and R.B. Wiringa, "Quantum Monte Carlo calculations of nuclei with A$\leq$7", 
               Phys. Rev. C {\bf 56}, 1720  (1997)
%
\bibitem{Eea02} E. Epelbaum  {\it et al.},  Phys. Rev. C {\bf 66}, "Three-nucleon forces from chiral effective field theory",  064001  (2002)
%
\bibitem{GP06} A. Gardestig and D.R. Phillips,"How low-energy weak reactions can constrain three-nucleon forces and the neutron-neutron scattering length",  Phys. Rev. Lett. {\bf 96}, 232301 (2006)
%
\bibitem{GQN09} D. Gazit, S. Quaglioni, and P.  Navr\'atil, "Three-Nucleon Low-Energy Constants from the Consistency of Interactions and Currents in Chiral Effective Field Theory",
  Phys. Rev. Lett. {\bf 103}, 102502 (2009)
%
\bibitem{Mea18}  L. E. Marcucci, F. Sammarruca, M. Viviani, and
  R. Machleidt, "Momentum distributions and short-range correlations in the deuteron and 3He with modern chiral potentials",  Phys. Rev. C {\bf 99}, 034003 (2019)
%
\bibitem{nu}  M. Abdullah {\it et al.}, ''Coherent elastic neutrino-nucleus scattering: Terrestrial and
astrophysical applications'', {\tt arXiv:2203:07361}
%
\bibitem{Pastore09}S. Pastore, L. Girlanda, R. Schiavilla, M. Viviani, and R. B. Wiringa,  ''Electromagnetic Currents and Magnetic Moments in $\chi$EFT'',  Phys. Rev. C {\bf 80}, 034004 (2009)

\bibitem{Shen12} G. Shen, L. E. Marcucci, J. Carlson, S. Gandolfi, and R. Schiavilla, ''Inclusive neutrino scattering off deuteron from threshold to GeV energies'',  Phys. Rev. C {\bf 86}, 035503 (2012) 
%
\bibitem{Newstead21} J. L. Newstead, R. F. Lang, and L. E. Strigari, ''Atmospheric neutrinos in next-generation xenon and argon dark matter experiments'',  Phys. Rev. D {\bf 104},  115022 (2021)
%
\bibitem{PanciPC} P. Panci, private communication


%


%



\end{thebibliography}
\end{document}